\documentclass[11pt]{article}
\usepackage{amssymb}
\usepackage{amsmath}
\usepackage{amstext}
\usepackage{graphicx,epsfig}
\usepackage{epsfig}
\usepackage{verbatim} 
\usepackage{fancybox}
\usepackage{color}
\usepackage{ulem}
\usepackage{enumitem}
\usepackage{subfigure}
\usepackage{bbm}
\usepackage{parskip}
\usepackage{dsfont}
\usepackage[numbers,sort&compress]{natbib}
\usepackage{ifthen}

\newcommand{\Comment}[1]{{}}
\definecolor{darkblue}{rgb}{0.15,0.35,0.55}
\definecolor{reddish}{rgb}{0.65, 0.2, 0.2}
\definecolor{forestgreen}{rgb}{0.133,0.545,0.133}

\usepackage[linktocpage=true]{hyperref}
\hypersetup{
colorlinks=true,
citecolor=darkblue,
linkcolor=reddish,
urlcolor=darkblue,
pdfauthor={},
pdftitle={},
pdfsubject={}
}

\makeatletter
\renewcommand{\paragraph}{%
  \@startsection{paragraph}{4}%
  {\z@}{.5ex \@plus 1ex \@minus .2ex}{-1em}%
  {\normalfont\normalsize\bfseries}%
}
\makeatother

\newcommand{\be}{\begin{equation}}
\newcommand{\ee}{\end{equation}}
\newcommand{\bea}{\begin{eqnarray}}
\newcommand{\eea}{\end{eqnarray}}
\newcommand{\beas}{\begin{eqnarray*}}
\newcommand{\eeas}{\end{eqnarray*}}

\newcommand{\Lstar}{\Lambda_{\star}}
\newcommand{\Lv}{\Lambda_{\rm v}}

\newcommand{\rstar}{r_{\star}}
\newcommand{\rv}{r_{\rm v}}

\newcommand{\p}{\partial}
\def\({\left(}
\def\){\right)}

\newcommand{\rd}{{\rm d}}
\newcommand{\vp}{\varphi}
\newcommand{\mpl}{M_{\rm Pl}}

\def\gsim{ \lower .75ex \hbox{$\sim$} \llap{\raise .27ex \hbox{$>$}} }
\def\lsim{ \lower .75ex \hbox{$\sim$} \llap{\raise .27ex \hbox{$<$}} }

\newboolean{editorial}
\setboolean{editorial}{true}
\newcommand{\editorial}[2]{\ifthenelse{\boolean{editorial}}{\textcolor{forestgreen}{ [\textsf{\textbf{{#1}}}:} \textcolor{darkblue}{\textsf{{#2}}}\textcolor{darkblue}{]}}{}}

\title{}
\author{}

\begin{document}
%
%\maketitle
\renewcommand{\thefootnote}{\fnsymbol{footnote}}
~
\vspace{2.5truecm}
\begin{center}
{\huge \bf{Double screening}}
\end{center} 

\vspace{1truecm}
\thispagestyle{empty}
\centerline{{\Large Pierre Gratia,${}^{\rm a,}$\footnote{\color{darkblue}{{\tt pgratia@uchicago.edu}}} Wayne Hu,${}^{\rm b,c,}$\footnote{\color{darkblue}{{\tt whu@background.uchicago.edu}}} Austin Joyce${}^{\rm c,}$\footnote{\color{darkblue}{{\tt ajoy@uchicago.edu}}} and Raquel H. Ribeiro${}^{\rm d,}$\footnote{\color{darkblue}{{\tt r.ribeiro@qmul.ac.uk}}}}}
\vspace{.7cm}

\centerline{\it ${}^{\rm a}$Department of Physics, University of Chicago, Chicago, IL 60637}

\vspace{.2cm}

\centerline{\it ${}^{\rm b}$Department of Astronomy and Astrophysics, University of Chicago, Chicago, IL 60637}

\vspace{.2cm}

\centerline{\it ${}^{\rm c}$Enrico Fermi Institute and Kavli Institute for Cosmological Physics,}
\centerline{\it University of Chicago, Chicago, IL 60637}

\vspace{.2cm}

\centerline{\it ${}^{\rm d}$School of Physics and Astronomy, Queen Mary University of London,}
\centerline{\it Mile End Road, London, E1 4NS, U.K.}

 \vspace{.8cm}
\begin{abstract}
\noindent
Attempts to modify gravity in the infrared typically require a screening mechanism to ensure consistency with local tests of gravity. These screening mechanisms fit into three broad classes; we investigate theories which are capable of exhibiting more than one type of screening. Specifically, we focus on a simple model which exhibits both Vainshtein and kinetic screening. We point out that due to the two characteristic length scales in the problem, the type of screening that dominates depends on the mass of the sourcing object, allowing for different phenomenology at different scales. We consider embedding this double screening phenomenology in a broader cosmological scenario and show that the simplest examples that exhibit double screening are radiatively stable. 
\end{abstract}

\newpage

\setcounter{tocdepth}{2}
\tableofcontents
\newpage
\renewcommand*{\thefootnote}{\arabic{footnote}}
\setcounter{footnote}{0}

\section{Introduction}
The observed accelerated expansion of the universe~\cite{Riess:1998cb,Perlmutter:1998np}---combined with the naturalness problems of the simplest explanation in terms of a cosmological constant~\cite{Weinberg:1988cp}---has catalyzed a reconsideration of gravitational physics in the infrared. Many different models have been considered, but essentially all the proposals for new physics in the gravitational sector introduce additional degrees of freedom, which most often are scalars~\cite{Clifton:2011jh, Joyce:2014kja}. Typically these new degrees of freedom are light, with masses of order the Hubble scale today, $m\sim H_0$, and couple to visible matter. Consequently, they mediate a force between matter sources. In order for this to not conflict with local tests of gravity, infrared modifications of gravity must therefore employ a {\it screening mechanism} to suppress their effects at small scales.

Various mechanisms have been proposed which allow these theories to evade local tests of gravity; they can be classified according to the following scheme~\cite{Joyce:2014kja}:
\begin{itemize}

\item Screening by field nonlinearities: In these models, screening activates in regions of large Newtonian potential, typically inducing large nonlinearities in the field itself which causes the fifth force to shut off. Examples of screening mechanisms which fall into this category are chameleon~\cite{Khoury:2003rn}, symmetron~\cite{Hinterbichler:2010es} and dilaton~\cite{Damour:1994zq} screening.

\item Screening with large first derivatives: In these models, screening occurs when the first gradient of the Newtonian potential becomes large, corresponding to large local gravitational acceleration. This causes the first derivatives of the field itself to become large and diminish the fifth force. This mechanism is employed in $k$-mouflage theories~\cite{Babichev:2009ee}, $P(X)$~\cite{Brax:2012jr, deRham:2014wfa} and DBI-type models~\cite{Burrage:2014uwa}; as well as in some attempts to embed MOND~\cite{Milgrom:1983ca} phenomenology in a cosmological framework~\cite{Babichev:2011kq,Khoury:2014tka}. We will refer to screening of this type as {\it kinetic screening}.

\item Screening by large second derivatives: Finally, screening can occur in regions where the second derivative of the Newtonian potential is large, corresponding to regions of high density. In these models, second gradients of the field become large, causing the fifth force to be screened. Screening in this class typically goes under the name the {\it Vainshtein mechanism}~\cite{Vainshtein:1972sx}. This is the screening mechanism which operates in massive gravity~\cite{Vainshtein:1972sx,Deffayet:2001uk,Chkareuli:2011te}, in DGP~\cite{Dvali:2000hr} and in the galileon~\cite{Nicolis:2008in}.
\end{itemize}
This classification is sometimes presented as being exclusive: {\it i.e.}, a given model can belong to only one category. In this paper we reconsider this and explore models where more than one type of screening can be active. Specifically, we focus on models which exhibit both kinetic and Vainshtein screening in different regimes. The simplest model in this class is that of a scalar field described by the lagrangian
\be
{\cal L} = -\frac{1}{2}(\partial\phi)^2 - \frac{c_3}{\Lambda_{\rm v}^3}\square\phi(\partial\phi)^2-\frac{c_4}{4\Lambda_{\star}^4}(\partial\phi)^4,
\label{X2plusgal}
\ee
where $c_i = \pm 1$.
We will mostly focus on the phenomenology of this particular model, but the general philosophy is much broader.\footnote{Other models with a similar derivative structure include the conformal galileon~\cite{Nicolis:2008in}
\be
{\cal L} = -\frac{f^2}{2}e^{2\pi}(\partial\pi)^2  - \frac{f^3}{\Lambda^3}\left(2\square\pi(\partial\pi)^2+(\partial\pi)^2\right)
\ee
and the cubic DBI galileon~\cite{deRham:2010eu}
\be
{\cal L} = \Lambda_\star^4\sqrt{1-\frac{(\partial\phi)^2}{\Lambda_\star^4}}+ \frac{c_3}{\Lambda_{\rm v}^3}\partial_\mu\phi\partial^\mu\partial^\nu\phi\partial_\nu\phi \left(1-\frac{(\partial\phi)^2}{\Lambda_\star^4}\right)^{-1},
\ee
with some signs flipped to allow for kinetic screening~\cite{Burrage:2014uwa}. Note that in contrast to the model~\eqref{X2plusgal}---which has an approximate galileon symmetry---the conformal and DBI models have exact symmetries: $\delta\pi = 2x^\mu+\left(2x^\mu x^\nu\partial_\nu-x^2\partial^\mu\right)\pi$ and $\delta\phi = x^\mu-\Lambda_\star^{-4}\phi\partial^\mu\phi$. These models have been investigated in numerous contexts, and it would be interesting to explore the double screening phenomenology in these cases.
}
 The essential new phenomenon is that the theory~\eqref{X2plusgal} has two characteristic scales in the presence of a massive source: the kinetic screening radius, $r_\star \propto \Lambda_\star^{-1}M^{1/2}$ and the Vainshtein radius, $r_{\rm v} \propto \Lambda_{\rm v}^{-1}M^{1/3}$, which scale {\it differently} with mass. We therefore see that for light objects, Vainshtein screening can be dominant, while for heavy objects kinetic screening can be most important.\footnote{We will be more precise later about what we mean by ``light" and ``heavy" objects.}
 
 The fact that kinetic screening is dominant on the largest scales is quite interesting, as it has been argued that kinetic screening is somewhat inefficient on quasi-linear scales~\cite{Brax:2014yla}, allowing for ${\cal O}(1)$ departures from $\Lambda$CDM phenomenology. The double screening mechanism allows for such interesting physics on cluster scales, while easily satisfying local tests of gravity due to the presence of the galileon term. Another intriguing aspect of the kinetic screening operators is that they are {\it not} invariant under adding constant gradient long-wavelength modes, opening the possibility of apparent violations of the equivalence principle on the largest scales~\cite{Creminelli:2013nua}.

In the lagrangian~\eqref{X2plusgal}, the galileon and $(\partial\phi)^4$ terms are suppressed by different scales $\Lambda_{\rm v}$ and $\Lambda_\star$; we will focus on the region of parameter space where $\Lambda_\star \gg \Lambda_{\rm v}$, primarily because this hierarchy of scales is radiatively stable. Terms which are quantum-mechanically generated are suppressed by powers of $\Lambda_{\rm v}/\Lambda_\star$, essentially because of the enhanced galileon symmetry in the limit where $\Lambda_\star \to \infty$~\cite{Pirtskhalava:2015nla}. This model is not the unique action with this property, but can be considered a representative example of this class which exhibits the double screening behavior we are interested in. Much of what we say can be extended to the more general case, and we will comment on these generalizations.

A somewhat similar setup was considered in~\cite{Babichev:2011kq}, consisting of a galileon plus a term of the form ${\cal L} \sim -(\partial\phi)^2\sqrt{\lvert (\partial\phi)^2\rvert}/(a_0M_{\rm Pl})$, in order to reproduce the MOND~\cite{Milgrom:1983ca} force law in galaxies---with $a_0$ the MOND crossover acceleration---while being screened at solar system scales. Like the case we consider, this theory possesses two regimes, where the force law is qualitatively different. However, our motivations are different; we imagine that $\phi$ has something to do with dark energy rather than dark matter phenomenology, and we emphasize that the length scales in the problem are mass-dependent, such that the observed phenomenology depends on the masses of the objects involved.

We begin in Section~\ref{introtoscreening} by considering the theory~\eqref{X2plusgal} on flat space in the presence of a massive source, and explore how the type of screening that occurs depends both on the mass of the source and the ratio of scales $\Lambda_{\rm v}/\Lambda_\star$. Interestingly, we find that in order for a real solution to exist for an arbitrary mass source, it is necessary to choose $c_3 = c_4=1$ in~\eqref{X2plusgal}; these are the choices of signs for these terms which would cause them to screen without the other present. We also consider generalizations beyond~\eqref{X2plusgal} to general $P(X)$ theories. One reason for considering this type of double screening is to alleviate the superluminality issues which plague Vainshtein and kinetic screening individually. However, we will show that superluminalities persist even in the presence of both terms.

In Section~\ref{sec:pheno}, we consider embedding double screening into a complete cosmology. There are cosmological solutions involving both the galileon and the $(\partial\phi)^4$ term which self-accelerate at the background level. This is a cosmological analogue of ``double screening," where the physics of linear perturbations is governed by a different operator than that which controls nonlinear physics. The phenomenological signatures of this scenario are somewhat similar to the ones discussed above. However, we will show that the parameters must be chosen in such a way that collapsed objects cannot exhibit the double screening phenomenon discussed in Section~\ref{introtoscreening}. We show that by introducing a cosmological constant (CC), it is possible to find a satisfactory cosmology where the $P(X)$ part of the lagrangian drives the background and which exhibits double screening for collapsed objects. This provides a proof-of-concept that the double screening phenomenology is compatible with cosmology.

We then turn to discuss the radiative stability of the theory in Section~\ref{sec:rad}; we give a simple power counting argument that the theory is quantum-mechanically stable---even trusting power law divergences---below the scale $\Lambda_\star$. However, since kinetic screening manifestly relies on trusting the theory above the scale $\Lambda_\star$, we then compute the 1-loop effective action and show that the generated terms are subdominant even above this scale so long as we only trust the logarithmic divergences.

Finally, we discuss general lessons which may be abstracted from our analysis and comment on natural future directions to explore in Section~\ref{sec:summary}. 

\paragraph{Conventions:} We work in mostly plus signature throughout, and define the reduced Planck mass by $\mpl^{-2}\equiv 8\pi G$.

\section{Double screening around point sources}
\label{introtoscreening}
As a first step toward exploring the combined effects of kinetic and Vainshtein screening, we consider the theory~\eqref{X2plusgal} on a fixed flat background and consider a coupling to matter of the form  $\frac{g \phi}{M_{\rm Pl}}T$. As mentioned before, the theory~\eqref{X2plusgal} consists of two distinct parts: a cubic galileon interaction and a $(\partial\phi)^4$-type interaction. We define the constants $c_3,c_4$ to take the values $\pm 1$; their magnitudes may be absorbed into the definitions of $\Lambda_\star,\Lambda_{\rm v}$. This lagrangian is exactly invariant under a shift symmetry,
$\phi\mapsto\phi+c.$
This symmetry is broken by the coupling to matter, but given that this breaking is $M_{\rm Pl}$ suppressed, it is relatively soft.
The motivation for considering this particular form of the action is that it is the simplest example of a theory which has operators that give rise to kinetic screening and Vainshtein screening. In Section~\ref{sec:px} we consider more general actions with similar phenomenology.

The equation of motion following from~\eqref{X2plusgal} can be written as
\be
\partial_\mu\left(\partial^\mu\phi+\frac{2c_3}{\Lambda_{\rm v}^3}\square\phi\partial^\mu\phi-\frac{c_3}{\Lambda_{\rm v}^3}\partial^\mu(\partial\phi)^2+\frac{c_4}{\Lambda_\star^4}(\partial\phi)^2\partial^\mu\phi\right) = - \frac{g}{M_{\rm Pl}}T~.
\label{eq:screeningpointsource}
\ee
Notice that the left-hand side takes the form $\partial_\mu J^\mu$; this is a consequence of the shift symmetry of the original lagrangian, $J^\mu$ is the Noether current associated to this shift symmetry. 

We solve this equation in the presence of a non-dynamical, massive, spherically-symmetric source with a stress tensor with trace $T = - M \delta^{(3)}(\vec x)$,
and search for static, spherically-symmetric profiles for the background field configuration ($\phi = \phi(r)$). With these simplifications, the equation of motion can be straightforwardly integrated to yield a polynomial equation for $\phi'$:
\be
\phi'+\frac{4c_3}{\Lambda_{\rm v}^3}\frac{\phi'^2}{r}+\frac{c_4}{\Lambda_\star^4}\phi'^3 = \frac{g M}{M_{\rm Pl}}\frac{1}{4\pi r^2}~.
\label{x2plusgalcubic}
\ee
This is a cubic equation, so it admits an exact solution in radicals, with three branches. Without loss of generality, we assume that  $g>0$, as its sign can be reabsorbed into a field redefinition of $\phi$.

The explicit solution to~\eqref{x2plusgalcubic} is somewhat cumbersome, so in order to develop some intuition for how this general solution will behave, 
we will first review some limits of the equation. In particular, we can first consider the limit $\Lambda_{\rm v}\to \infty$
to see how screening of the kinetic type operates~\cite{Dvali:2012zc,Brax:2012jr,Gabadadze:2012sm}. In this limit, the solution to~\eqref{x2plusgalcubic} has two asymptotic regimes: very far from the source and close to the object. These two regimes are separated by the characteristic length scale where $\phi' \sim \Lambda_\star^2$,
\be
r_\star  = \frac{1}{\Lambda_\star}\left(\frac{g M}{M_{\rm Pl}}\right)^{1/2},
\label{kradius}
\ee
which we will refer to as the {\it kinetic screening radius}. Far from the source, the kinetic self-interaction is irrelevant and the field has the expected Newtonian $\phi\sim 1/r$ profile. However, close to the source, the non-linear $(\partial\phi)^4$ term dominates so that the field profile in the asymptotic regimes is
\be
\phi'(r) \simeq \left\{\begin{array}{lr} \displaystyle{\frac{\Lambda_\star^2}{4\pi}\left(\frac{r_\star}{r}\right)^2}&~~~~~~~~~~~~~~~~~~{\rm for}~~~~~~r\gg r_\star \\ \\
\displaystyle{c_4^{1/3}\frac{\Lambda_\star^2}{(4\pi)^{1/3}} \left(\frac{r_\star}{r}\right)^{2/3}} &~~~~~~~~~~~{\rm for}~~~~~~r\ll r_\star \end{array} \right. \ .
 \label{phiprimeeqn}
\ee
Note that a continuous solution which exists at all $r$ is possible only if $c_4 = 1$.

The force of gravity due to an object of mass $M$ is given by
\be
\vec F_{\rm grav.} =\hat r \frac{M}{8\pi M_{\rm Pl}^2}\frac{1}{r^2} = \hat r\frac{\Lambda_\star^2}{8\pi g M_{\rm Pl}}\left(\frac{r_\star}{r}\right)^2\,,
\ee
so that the ratio of the so-called fifth force mediated by the scalar ($\vec F_\phi = \hat r\frac{g}{M_{\rm Pl}}\phi'$) to that of gravity close to the source is
\be
\frac{F_\phi}{F_{\rm grav.}} = 2g^2 \left(\frac{r}{r_\star}\right)^{4/3}~~~~~~{\rm for}~~~~~r\ll r_\star, 
\label{pxforceratio}
\ee
which goes to zero as we approach the source---this is {\it screening} and it explains why it is possible for light scalar fields to evade detection in the solar system, their contribution to tests of gravity is strongly suppressed.

Additionally, we can take the limit $\Lambda_\star \to \infty$ in~\eqref{x2plusgalcubic}, so that it reduces to the equation of motion of the galileon~\cite{Nicolis:2008in}. This theory exhibits Vainshtein screening~\cite{Vainshtein:1972sx,Deffayet:2001uk}. As before,
there are two regimes; in this case they are separated by a different length scale, known as the {\it Vainshtein radius}
\be
r_{\rm v} = \frac{1}{\Lambda_{\rm v}}\left(\frac{g M}{M_{\rm Pl}}\right)^{1/3}\ .
\label{vradius}
\ee
In order for a real solution to exist everywhere, we must have $c_3g > 0$, so we must take $c_3 = 1$ (note that we have defined $g > 0$). With these considerations the gradient of the field profile can be solved for directly
\be
\phi'(r) = \frac{r \Lambda_{\rm v}^3}{8}\left(-1+\sqrt{1+\frac{4}{\pi}\left(\frac{r_{\rm v}}{r}\right)^3}\right)\,,
\ee
and we can see that it also simplifies in the asymptotic regimes:
\be
\phi'(r) \simeq \left\{\begin{array}{lr} \displaystyle{\frac{\Lambda_{\rm v}^3r_{\rm v}}{4\pi} \left( \frac{r_{\rm v}}{r}\right)^2}&~~~~~~~~~~~~~~~~~~{\rm for}~~~~~~r\gg r_{\rm v} \\ \\
 \displaystyle{\frac{\Lambda_{\rm v}^3 r_{\rm v}}{(16\pi)^{1/2}}\left(\frac{r_{\rm v}}{r}\right)^{1/2}} &~~~~~~~~~~~{\rm for}~~~~~~r\ll r_{\rm v} \end{array} \right. \ .
\ee
As in the kinetic case, the ratio of the galileon force to that of gravity drops off sharply inside the Vainshtein radius, this time scaling as $F_\phi/F_{\rm grav.} \sim r^{3/2}$. Note that this goes to zero {\it faster} than~\eqref{pxforceratio} indicating more efficient screening compared to the kinetic case.

A reasonable question to ask is whether both of these mechanisms can operate simultaneously---or, more generally, if there can be any interesting interplay between the terms responsible for screening. In order to answer this question, we now investigate the solution of the equation~\eqref{x2plusgalcubic} in full generality.

\subsection{Combining the two mechanisms}
\label{sec:intuition}

As we have seen above, around a massive source there are now {\it two} interesting length scales, the kinetic screening radius, $\rstar$, given in~\eqref{kradius} and the Vainshtein radius, $r_{\rm v}$, in~\eqref{vradius}.\footnote{It is possible to generalize this notion to theories which screen via $N^{\rm th}$ derivatives becoming large: $\partial^N\phi \sim \Lambda^{N+1}$ which will happen around a point source at the characteristic radius 
\begin{equation*}
r_N \sim \frac{1}{\Lambda}\left(\frac{gM}{M_{\rm Pl}}\right)^\frac{1}{N+1}.
\end{equation*}
It is then possible to imagine a cascade of different screening behaviors where the field profile changes near each of these length scales. Since there are not any known explicit examples of theories which screen with third or higher derivatives (in Lorentz invariant theories, this is likely to imply ghost instabilities, though it might be possible in theories of the type considered in~\cite{Hinterbichler:2014cwa,Griffin:2014bta}), we will not explore this notion further presently. We thank Kurt Hinterbichler for pointing out this possibility.
}
Generically, either of these two length scales could be larger, depending upon how we choose the hierarchy between $\Lambda_\star$ and $\Lambda_{\rm v}$. However, as we will argue in Section~\ref{sec:rad}, a well-motivated corner of parameter space to explore is where there is a hierarchy
\be
\Lambda_\star \gg \Lambda_{\rm v},
\ee
owing to the radiative stability of this choice.
This stability can be thought of as a vestige of the ``weakly-broken" galileon invariance of the theory which is restored in the limit that $\Lambda_{\rm v}/\Lambda_\star \to 0$, as was noted in~\cite{Pirtskhalava:2015nla}. In Section~\ref{sec:rad}, we give a self-contained argument that this hierarchy is radiatively stable and show that even on backgrounds where both
\be
\frac{\square\phi}{\Lambda_{\rm v}^3}\gg 1,~~~~~~~~~~~~~~\frac{(\partial\phi)^2}{\Lambda_{\star}^4}\gg 1,
\ee
loop corrections are under control. This is quite intriguing because it allows us to simultaneously realize both screening mechanisms we have discussed.

In light of this nice quantum-mechanical behavior, we will focus on the situation where $\Lambda_\star \gg \Lambda_{\rm v}$ in what follows. In any case, relaxing this hierarchy between the two scales does not drastically change the coarse features of the phenomenology, though it does make the theory significantly more tuned.

Before explicitly solving the equation of motion following from~\eqref{X2plusgal}, it is first helpful to get some intuition for how the solutions will behave. Probably the most important difference between the present case and situations where only either Vainshtein screening or kinetic screening operate is the presence of two length scales which scale {\it differently} with the mass of the object sourcing the field profile. In particular, the kinetic screening radius grows more quickly with mass than the Vainshtein radius. 

\begin{figure}
\centering
\includegraphics[width=6in]{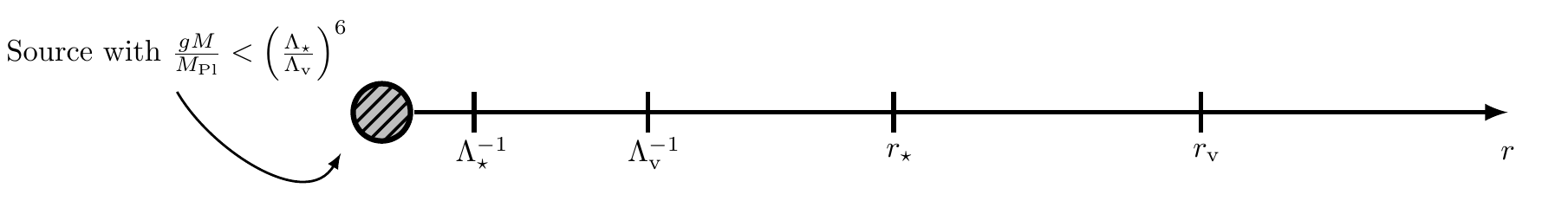}
\caption{\label{fig:rvbigger}\small Schematic of relevant scales, when $\Lambda_\star \gg \Lambda_{\rm v}$ and $gM/M_{\rm Pl} < (\Lambda_\star/\Lambda_{\rm v})^6$. In this case, the Vainshtein radius, $r_{\rm v}$, is the larger of the two scales, $r_{\rm v} > r_\star$. In this situation, the phenomenology is essentially the same as in the theory of the galileon alone. This diagram is not drawn to scale.}
\end{figure}

\begin{figure}
\centering
\includegraphics[width=6in]{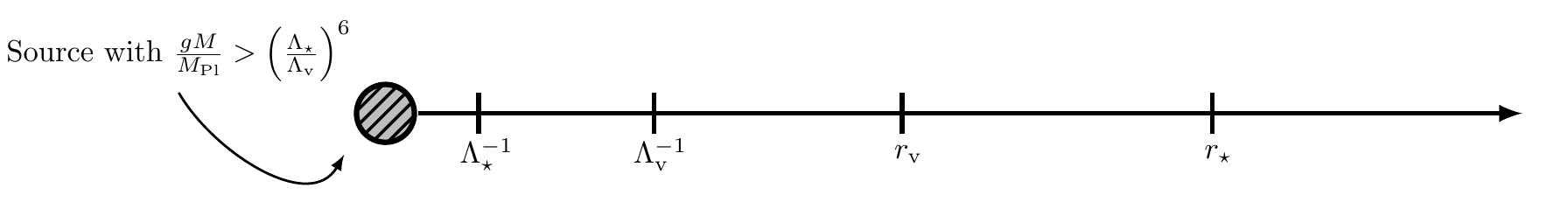}
\caption{\label{fig:rsbigger}\small Schematic of relevant scales, when $\Lambda_\star \gg \Lambda_{\rm v}$ and $gM/M_{\rm Pl} > (\Lambda_\star/\Lambda_{\rm v})^6$. In this case, the kinetic radius, $r_\star$, is the larger of the two scales, $r_\star > r_{\rm v}$. In this situation there are three distinct regimes, very far from the source the potential behaves like a normal $1/r$ law. As we approach the source, the force is initially kinetic screened. Even closer to the source, Vainshtein screening takes over. This diagram is not to scale.}
\end{figure}

This difference in scaling turns out to be essential to the phenomenology. In particular, since we are working in the regime $\Lambda_\star\gg\Lambda_{\rm v}$, small mass objects will have their $r_{\rm v} > r_\star$ (as is depicted in Figure~\ref{fig:rvbigger}), and they will experience screening in much the same way as in the galileon model, the $(\partial\phi)^4$ term will never become important. However, for large mass objects, the kinetic screening radius $r_\star$ can be the larger of the two radii (see Figure~\ref{fig:rsbigger}), so objects will first experience the nonlinear effects of the $(\partial\phi)^4$ and then at smaller radii the galileon term will come to dominate and close to the object the field profile will mimic that of the galileon. We can see that this crossover in behavior occurs where $r_\star \sim r_{\rm v}$, corresponding to the mass
\be
\frac{gM}{M_{\rm Pl}} \sim \left(\frac{\Lambda_\star}{\Lambda_{\rm v}}\right)^6,
\ee
which depends on the hierarchy of scales chosen to suppresses the two operators, as might be expected.

Concretely, we see that for sources satisfying
\be
\frac{gM}{M_{\rm Pl}} \ll \left(\frac{\Lambda_\star}{\Lambda_{\rm v}}\right)^6~,
\label{smallm}
\ee
we will have $r_{\rm v} > r_\star$, while in the opposite regime:
\be
\frac{gM}{M_{\rm Pl}} \gg \left(\frac{\Lambda_\star}{\Lambda_{\rm v}}\right)^6~,
\label{largem}
\ee
the kinetic screening radius will be larger, $r_\star > r_{\rm v}$. Therefore, matter sources with different mass $M$ will dictate a different hierarchy between the kinetic and Vainshtein radii. We now explore both of these regimes.

\subsubsection{Small mass sources}
If the source satisfies the inequality~\eqref{smallm}, as we approach the object from infinity, the field profile will initially be of the usual Newtonian $1/r$ form and any test objects will feel an un-screened fifth force mediated by $\phi$. Eventually, at the scale $r_{\rm v}$, the cubic galileon term will begin to dominate the dynamics, and the field profile will be of the approximate form:
\be
\phi'(r) \sim \Lambda_{\rm v}^3 r_{\rm v} \left(\frac{r_{\rm v}}{r}\right)^{1/2}~~~ \textrm{for} \ ~~ r\,\lsim\,\rv.
\label{galfieldprofile}
\ee
Now, the na\"ive expectation is that the field profile will be close to this form until one reaches $r_\star$, where it will then switch over to being of the form~\eqref{phiprimeeqn}. However, this intuition is incorrect, as can be seen by evaluating how both the galileon and $X^2$ operators scale on the background~\eqref{galfieldprofile}
\begin{align}
\label{galongalbkgrd}
\frac{1}{\Lambda_{\rm v}^3}\square\phi(\partial\phi)^2\Big\rvert_{\phi'\sim r^{-1/2}}&\sim \left(\Lambda_{\rm v}^3 r_{\rm v}\right)^2 \left(\frac{r_{\rm v}}{r}\right)^{5/2}\\
\frac{1}{\Lambda_\star^4}(\partial\phi)^4\Big\rvert_{\phi'\sim r^{-1/2}}  &\sim \left(\Lambda_{\rm v}^2r_{\rm v}\right)^4\left(\frac{\Lambda_{\rm v}}{\Lambda_\star}\right)^4 \left(\frac{r_{\rm v}}{r}\right)^2~.
\end{align}
This makes it clear that if the galileon operator comes to dominate first (that is, at a further distance from the source), the $(\partial\phi)^4$ operator will never be equally important, as it grows more slowly with decreasing $r$. Another way of seeing this behavior is to check that by the time one reaches $\rstar$ and the kinetic interaction starts correcting the kinetic term, the galileon operator has grown very large already, which in fact makes the contribution of the kinetic screening term to the dynamics entirely negligible.

\subsubsection{Large mass sources}
Next, we consider sources in the opposite regime, satisfying~\eqref{largem}.
In this regime, $r_\star > r_{\rm v}$, so as we approach the source from infinity, we will first enter the kinetic screening regime, where the field profile is approximately
\be
\phi'(r) \sim \Lambda_\star^2\left(\frac{r_\star}{r}\right)^{2/3}~.
\ee
Note that the galileon term continues to grow more quickly than the kinetic screening term when evaluated on this background:
\begin{align}
\label{galonkinbkgd}
\frac{1}{\Lambda_{\rm v}^3}\square\phi(\partial\phi)^2\Big\rvert_{\phi'\sim r^{-2/3}}&\sim \Lambda_\star^3r_\star^{-1} \left(\frac{\Lambda_\star}{\Lambda_{\rm v}}\right)^3  \left(\frac{r_\star}{r}\right)^3\\
\label{kinonkinbkgd}
\frac{1}{\Lambda_\star^4}(\partial\phi)^4\Big\rvert_{\phi'\sim r^{-2/3}}  &\sim \Lambda_\star^4 \left(\frac{r_\star}{r}\right)^{8/3}~,
\end{align}
so we should expect it to come to dominate at some radius, $\bar r$. This distance can be estimated by equating~\eqref{galonkinbkgd} and~\eqref{kinonkinbkgd} to obtain:
\be
\frac{\bar r}{r_\star} \sim \left(\frac{\Lambda_\star}{\Lambda_{\rm v}}\right)^9\left(\frac{g M}{M_{\rm Pl}}\right)^{-3/2}~.
\label{vainshteintakesover}
\ee
Note that this distance is generically {\it smaller} than the Vainshtein radius:
\be
\frac{\bar r}{r_{\rm v}} \sim \left(\frac{\Lambda_\star}{\Lambda_{\rm v}}\right)^8\left(\frac{g M}{M_{\rm Pl}}\right)^{-4/3} < 1~,
\ee
where to write the inequality we have used~\eqref{largem}. We conclude that for large mass sources there exist two different screening regimes with a clear separation between them. It is important to assess whether or not the EFT can be trusted near $\bar{r}$, and we address this question in Section~\ref{sec:rad}.

A final comment concerns the signs of the coefficients in front of the operators in~\eqref{X2plusgal}; it might be expected that combining the two operators might loosen some of the sign constraints discussed in Section~\ref{introtoscreening}. For example, we may now think that we can set $c_4 = -1$ but nevertheless have screening due to the galileon operator. However, as we have just seen, by choosing a sufficiently massive source, we can cause the $(\partial\phi)^4$ operator to become important first, and unless its sign is chosen appropriately, a solution which exhibits screening will not exist. A similar argument holds for the galileon term. We therefore see, surprisingly, that there is no additional sign freedom afforded to us in this more general situation---each of the operators must have the ``correct" sign for them to screen in isolation.  

\subsection{Explicit solution around a point source}

Now that we have an intuitive understanding of how solutions to the equation of motion of~\eqref{X2plusgal} should behave around an isolated point source, let us solve  explicitly the equation of motion and verify our intuition.

Though the exact solutions to~\eqref{x2plusgalcubic} are fairly complicated, it is straightforward to verify numerically that a solution which is everywhere real and has the correct asymptotic behavior ({\it i.e.}, that $\phi'\sim 1/r^2$ as $r\to \infty$) exists for all values of the mass, $M$, {\it only} if $c_3 =1, c_4 =1$. This is in accord with the intuitive arguments of the previous section. The explicit solution itself is not particularly enlightening, but we quote it here for completeness
\be
\phi'(r) = -\Lambda_\star^2\left[\frac{4}{3}\left(\frac{r_{\rm v}}{r_\star}\right)^3 \left(\frac{r_\star}{r}\right) + 2\pi^{1/3}\left(\frac{r}{r_\star}\right)\frac{1}{Q(r)}-\frac{32\pi^{1/3}}{3}\left(\frac{r_{\rm v}}{r_\star}\right)^6\left(\frac{r_\star}{r}\right)\frac{1}{Q(r)} - \frac{1}{6\pi^{1/3}}\left(\frac{r_\star}{r}\right)Q(r)\right],
\label{fullsolndgpx2}
\ee
where the quantity $Q(r)$ is given by:
\be
\begin{array}{l}
Q(r)\equiv \Bigg[27\frac{r}{r_\star}+144\pi\left(\frac{r_{\rm v}}{r_\star}\right)^3\left(\frac{r}{r_\star}\right)^2-512\pi\left(\frac{r_{\rm v}}{r_\star}\right)^9\\
~~~~+3\sqrt{192\pi^2\left(\frac{r}{r_\star}\right)^6+81\left(\frac{r}{r_\star}\right)^2+864\pi\left(\frac{r_{\rm v}}{r_\star}\right)^3\left(\frac{r}{r_\star}\right)^3-768\pi^2\left(\frac{r_{\rm v}}{r_\star}\right)^6\left(\frac{r}{r_\star}\right)^4-3072\pi\left(\frac{r_{\rm v}}{r_\star}\right)^{9}\left(\frac{r}{r_\star}\right)}
~~\Bigg]^{1/3}.
\end{array}
\ee
Although $Q$ is not manifestly real (as the various roots can become imaginary), the combination~\eqref{fullsolndgpx2} is real for all values of the parameters.

\begin{figure}
\centering
\includegraphics[width=3.2in]{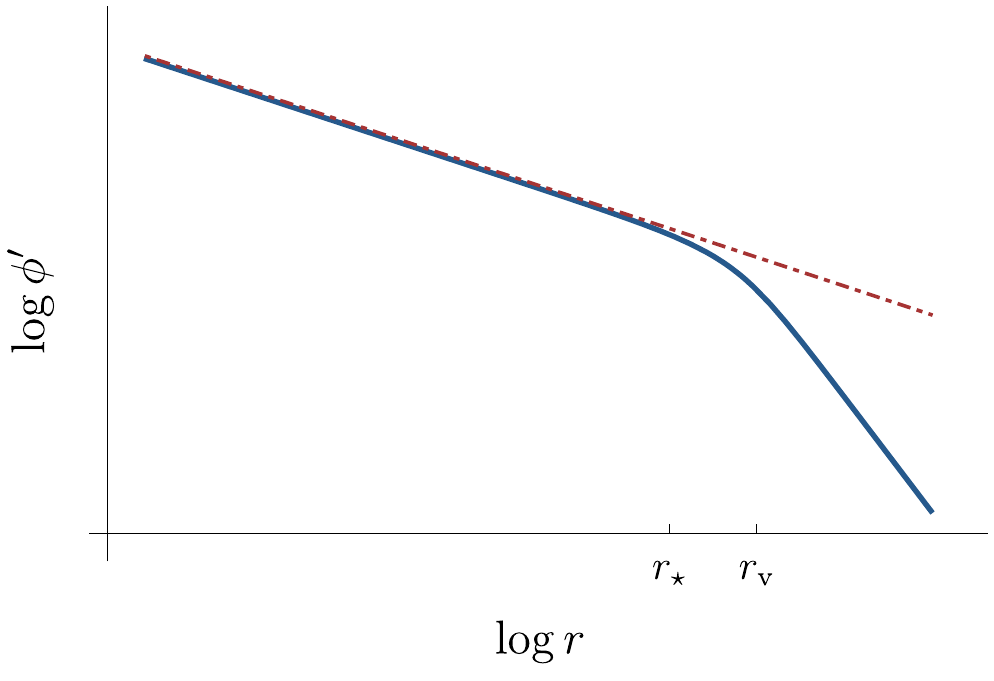}
\includegraphics[width=3.2in]{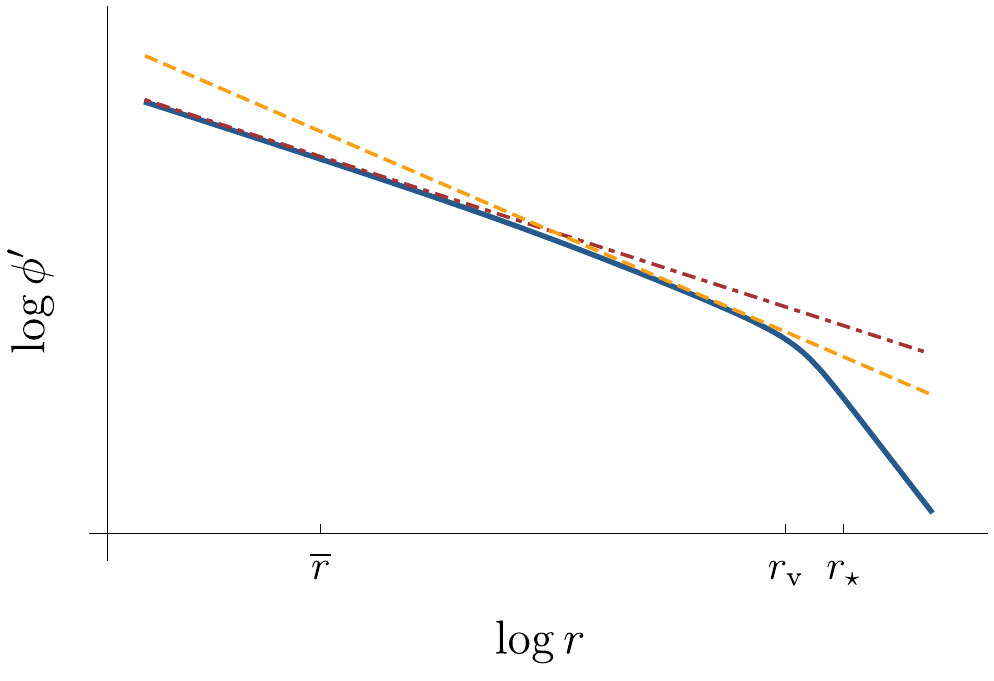}
\caption{\small\label{fig:solns}Here we plot the solution~\eqref{fullsolndgpx2} for two fiducial choices of parameters on a log-log scale. {\it Left}: Situation where the Vainshtein radius is larger than the kinetic screening radius (small mass objects: those satisfying~\eqref{smallm}). The parameters we have chosen are $M = 10^2M_{\rm Pl}, \Lambda_{\rm v} = 10^{-2}M_{\rm Pl}, \Lambda_\star = 10^{-1}M_{\rm Pl}, g=1$.
At large distances, the field profile scales as $r^{-2}$ while inside the Vainshtein radius, the profile scales as $r^{-1/2}$ all the way until $r = 0$, as expected. The red line (dot dashed) has slope $-1/2$ and is intended to guide the eye. {\it Right}: Situation where the kinetic screening radius is larger than the Vainshtein radius. Parameters are $M = 10^{10}M_{\rm Pl}, \Lambda_{\rm v} = 10^{-2}M_{\rm Pl}, \Lambda_\star = 10^{-1}M_{\rm Pl}, g=1$. Here we see that as the source is approached, the field first scales as $r^{-2/3}$ and then transitions to scaling as $r^{-1/2}$ at smaller radii. The red line is again of slope $-1/2$ while the orange line (dashed) has slope $-2/3$.}
\end{figure}

Note that~\eqref{fullsolndgpx2} has the properties expected of the field profile in Section~\eqref{sec:intuition}; in the limit $r\to\infty$, the solution takes the expected $1/r^2$ form:
\be
\phi'(r) \simeq \frac{gM}{4\pi M_{\rm Pl}}\frac{1}{r^2}\,,~~~~~{\rm as}~r\to\infty,
\ee
while in the opposite limit---close to the source---the field profile is that of the galileon, regardless of the mass of the object:
\be
\phi'(r) \simeq \frac{\Lambda_{\rm v}^3 r_{\rm v}}{(16\pi)^{1/2}}\left(\frac{r_{\rm v}}{r}\right)^{1/2},~~~~~{\rm as}~r\to0.
\ee
Further, in the limit that the kinetic screening radius is much larger than the Vainshtein radius ($r_{\rm v} \to 0$), we recover the correct solution for kinetic screening:
\be
\lim_{r_{\rm v}\to 0}\phi'(r) = -\frac{\Lambda_\star^2\left(\frac{8\pi}{3}\right)^{1/3} \left(\frac{r}{r_\star}\right)^{2/3}}{\left(9+\sqrt{81+192\pi^2\left(\frac{r}{r_\star}\right)^4}\right)^{1/3}} + \Lambda_\star^2(72\pi)^{-\frac{1}{3}}\left(9+\sqrt{81+192\pi^2\left(\frac{r}{r_\star}\right)^4}\right)^{1/3}\left(\frac{r_\star}{r}\right)^{2/3} \ .
\ee
And similarly in the limit that the kinetic screening radius is very small compared to the Vainshtein radius, we recover the galileon solution:
\be
\lim_{r_\star\to0}\phi'(r)= \frac{r \Lambda_{\rm v}^3}{8}\left(-1+\sqrt{1+\frac{4}{\pi}\left(\frac{r_{\rm v}}{r}\right)^3}\right)~.
\ee

It can also be seen that the solution~\eqref{fullsolndgpx2} exhibits the expected behaviors for small and large mass sources. For light sources, satisfying~\eqref{smallm}, the profile at large $r$ goes as $1/r^2$ before scaling as $1/\sqrt{r}$ as one approaches the source. In the opposite regime, where the source satisfies~\eqref{largem}, the profile at infinity is $1/r^2$ but now as we approach the source, the profile first scales as $1/r^{2/3}$ before turning over further and scaling as $1/\sqrt r$ at smaller radii. We plot these two situations for a fiducial choice of parameters in Figure~\ref{fig:solns}.

\subsection{Superluminality}
Superluminal propagation of fluctuations is extremely common in the field theories which arise from infrared modifications of gravity~\cite{Adams:2006sv}. Superluminality has been shown to occur in DGP~\cite{Hinterbichler:2009kq}, galileon~\cite{Nicolis:2009qm,Goon:2010xh} and $k$-essence~\cite{Babichev:2007dw} models; as well as in massive gravity~\cite{Deser:2012qx,Deser:2015wta,Motloch:2015gta}.\footnote{However, the spherically symmetric configurations which exhibit superluminality in galileon models {\it do not} in massive gravity~\cite{Berezhiani:2013dw}.} It is still somewhat unclear to what extent superluminality in an EFT implies a pathology~\cite{Dubovsky:2007ac,Burrage:2011cr,deRham:2014lqa}, but nevertheless it would be preferable for it to be absent. One motivation for adding the $(\partial\phi)^4$ operator to the galileon is to try to remove the superluminality present in the model. However, as we have seen above, we have to choose the sign of this operator such that this isn't possible.

\begin{figure}
\centering
\includegraphics[width=3.2in]{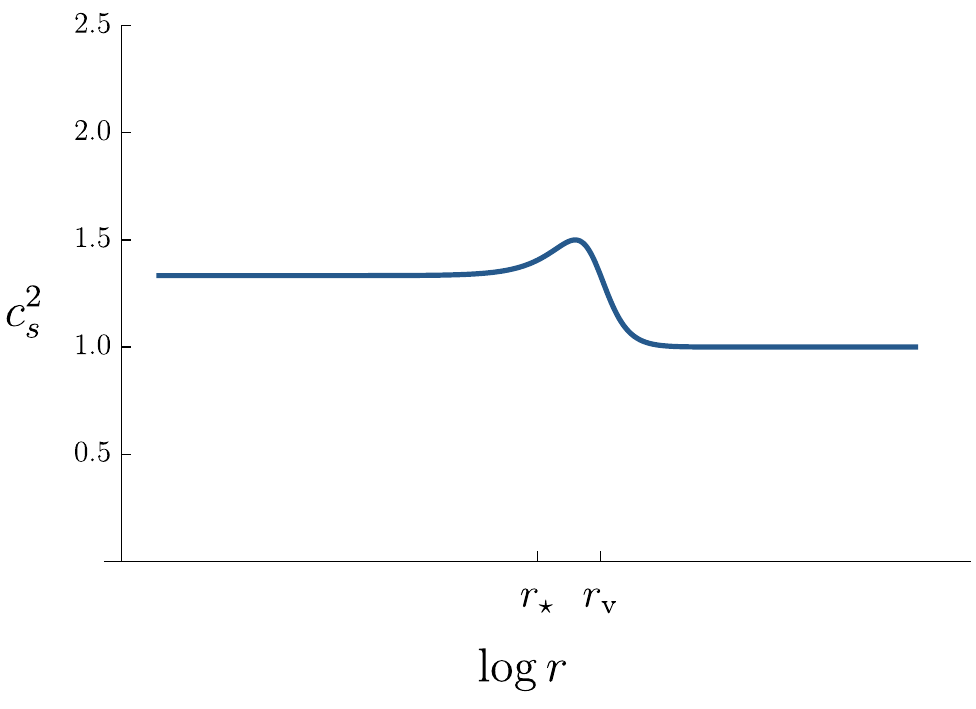}
\includegraphics[width=3.2in]{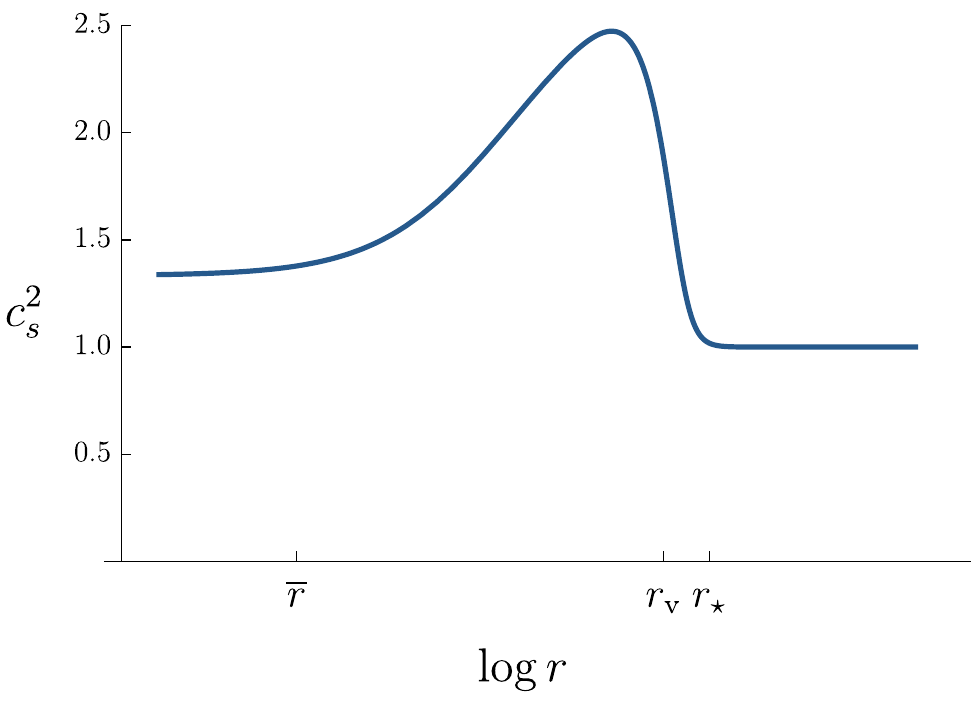}
\caption{\label{fig:cs}\small Log-linear plot of radial sound speed for perturbations about the solution~\eqref{fullsolndgpx2}. {\it Left}: Situation where $r_{\rm v} > r_\star$. The sound speed here is essentially the same as that of the galileon in isolation. Parameters chosen are the same as above: $M = 10^2M_{\rm Pl}, \Lambda_{\rm v} = 10^{-2}M_{\rm Pl}, \Lambda_\star = 10^{-1}M_{\rm Pl}, g=1$. {\it Right}: Radial sound speed where $r_\star > r_{\rm v}$. Here we see that the sound speed is similar to that in the $X+X^2$ case inside $r_\star$, where the sound speed is somewhat larger than when the galileon operator dominates. Inside the radius $\bar r$, where the galileon term becomes important, the sound speed approaches the same value as in the left panel.
Parameters here are $M = 10^{10}M_{\rm Pl}, \Lambda_{\rm v} = 10^{-2}M_{\rm Pl}, \Lambda_\star = 10^{-1}M_{\rm Pl}, g=1$.}
\end{figure}

In order to diagnose superluminality, we perturb the lagrangian~\eqref{X2plusgal}---setting $c_3=c_4=1$---as $\phi = \bar\phi + \varphi$, leading to the quadratic action for $\varphi$:
\be
{\cal L}_\vp = -\frac{1}{2}\bigg(\Big[1+\frac{4}{\Lambda_{\rm v}^3}\square\bar\phi+\frac{1}{\Lambda_\star^4}(\partial\bar\phi)^2\Big]\eta^{\mu\nu}-\frac{4}{\Lambda_{\rm v}^3}\partial^\mu\partial^\nu\bar\phi+\frac{2}{\Lambda_\star^4}\partial^\mu\bar\phi\partial^\nu\bar\phi\bigg)\partial_\mu\vp\partial_\nu\vp~,
\label{pertlag}
\ee
which reduces to the following for a spherically-symmetric background $\bar\phi = \bar\phi(r)$
\be
{\cal L}_\vp \equiv -\dfrac{1}{2}Z^{\mu\nu}\p_{\mu}\vp \p_{\nu}\vp= \frac{1}{2}\left(1+\frac{4}{\Lambda_{\rm v}^3}\left(\bar\phi''+\frac{2}{r}\bar\phi'\right)+\frac{1}{\Lambda_\star^4}\bar\phi'^2\right)\dot\vp^2 - \frac{1}{2}\left(1+\frac{8}{\Lambda_{\rm v}^3}\frac{\bar\phi'}{r}+\frac{3}{\Lambda_\star^4}\bar\phi'^2\right)(\partial_r\vp)^2 + (\partial_\Omega\vp)^2~,
\ee
where we have been schematic about the angular derivatives, and $Z^{\mu\nu}$ is commonly referred to as the kinetic matrix. The radial sound speed is then given by the expression
\be
c_s^2(r)\equiv \dfrac{Z^{rr}}{Z^{tt}} = \displaystyle{\frac{1+\frac{8}{\Lambda_{\rm v}^3}\frac{\bar\phi'}{r}+\frac{3}{\Lambda_\star^4}\bar\phi'^2}{1+\frac{4}{\Lambda_{\rm v}^3}\left(\bar\phi''+\frac{2}{r}\bar\phi'\right)+\frac{1}{\Lambda_\star^4}\bar\phi'^2}}~.
\ee
Since the eigenvalues of the kinetic matrix depend on the background radial profile,
 $\bar{\phi}(r)$, the resulting radial phase velocity of the fluctuations about this background will be different from unity. 
Indeed, it will become superluminal. We plot the radial sound speed for the two situations $r_{\rm v} > r_\star$ and $r_{\rm v}< r_\star$ in Figure~\ref{fig:cs}. The sound speed is everywhere greater than 1; inside the screening radius $c_s^2$ is ${\cal O}(1)$ greater than unity, while it asymptotes to 1 as $r\to\infty$.

More generally, it appears that superluminality comes together with kinetic or Vainshtein
screening. All known examples of theories which exhibit such screening also have superluminal fluctuations. As of yet, it has not been shown that such screening necessarily implies superluminality, but it is plausible that the requirements of screening (large eigenvalues of the kinetic matrix) generically lead to superluminality (the spatial eigenvalue being larger than the temporal one). It would be very interesting to show that this is always true.\footnote{Note that this can be shown for $P(X)$ theories (see {\it e.g.}, Section 6.1.4 of~\cite{Joyce:2014kja}.)}

\subsection{Generalization to $P(X)$}
\label{sec:px}
So far we have focused on the simplest model~\eqref{X2plusgal} which exhibits double screening. We now consider generalizing this model by allowing for a broader class of
 operators which kinetically screen.\footnote{A complementary generalization---which would be interesting to investigate---is to include higher-order galileon operators in the action. These will still have characteristic scale $r_{\rm v}$, but can behave differently from the cubic galileon.} To do this, we generalize~\eqref{X2plusgal} to contain an arbitrary function of $X \equiv -\frac{1}{2\Lambda_\star^4} (\partial\phi)^2$ as
\be
{\cal L} = \Lambda_\star^4P(X) - \frac{c_3}{\Lambda_{\rm v}^3}\square\phi(\partial\phi)^2.
\label{eq:galileonandgeneralpx}
\ee
Provided $P(X)$ is analytic, we may think of it as a Taylor series:\footnote{It is also interesting to consider cases where this breaks down and $n$ is not an integer~\cite{Babichev:2011kq,Khoury:2014tka}.}
\be
{\cal L}  = \Lambda_\star^4\sum_{n=1}^\infty c_n X^n - \frac{c_3}{\Lambda_{\rm v}^3}\square\phi(\partial\phi)^2.
\ee
We assume that the coefficients $c_n$ are chosen so that there exists a solution which screens at small radius which also asymptotes to $\phi \sim 1/r$ at large $r$.\footnote{In the pure $P(X)$ case, the requirement for such a solution to exist is for the quantity $P_{,X} > 0$ to be increasing as $-X$ increases from $0$ to $\infty$. In the $P(X)$ + galileon case, the constraints are the same, because we can find an arrangement of sources for which the $P(X)$ terms become important at a larger radius. Note that for cosmological solutions, $X > 0$, and for these solutions to exist, it must also be possible to invert the equation of motion on $(0,\infty)$, which imposes further constraints on the $c_n$. See~\cite{Barreira:2015aea} for a complete discussion of the constraints placed on the functional form of $P(X)$.} This is the analogue of having to choose $c_4 = 1$ in~\eqref{X2plusgal}. Notice that the choice of coefficients, $c_n$, is a technically natural one if we focus on radiative corrections independent of the regularization procedure, regardless of there being a finite or infinite number of operators in the lagrangian above---see Section~\ref{sec:rad} for more details. 

The radius where $|X|\sim 1$ again defines the kinetic screening radius, $r_\star$, which is the same in this case as in the simpler model:~\eqref{kradius}. Therefore, we see that there is again a separation between objects for which the Vainshtein radius is larger and those for which the kinetic screening radius is larger. 
First, note that when $X$ becomes large, so that the fifth force is screened, the term $X^n$ with the highest power will dominate and the field profile scales as
\be
\phi'(r) \sim \Lambda_\star^2\left(\frac{r_\star}{r}\right)^\frac{2}{2n-1}~~~~~~{\rm for}~~~r\ll r_\star~.
\label{xnprofile}
\ee

The phenomenology in this situation can be much different from the $X^2$ case, because operators of the form $X^n$ for $n > 2$ can grow faster than the galileon term. We first consider what happens for light sources where $r_{\rm v} > r_\star$. As we approach from infinity, the field profile first scales like that of the galileon:
\be
\phi'(r)\sim\Lambda_{\rm v}^3 r_{\rm v}\left(\frac{r_{\rm v}}{r}\right)^{1/2}~.
\ee
Note that on this background, the operator $X^n$ scales as:
\be
\Lambda_\star^4X^n\big\rvert_{\phi'\sim r^{-1/2}} \sim \Lambda_{\rm v}^{2n+4}\left(\frac{\Lambda_{\rm v}}{\Lambda_\star}\right)^{4n-4}r_{\rm v}^{2n}\left(\frac{r_{\rm v}}{r}\right)^n~,
\ee
while the galileon operator scales as~\eqref{galongalbkgrd}, which goes as $\sim r^{-5/2}$. In the case that we studied above, where $n=2$, the $X^2$ operator never comes to dominate, but we see that here if we take $n \geq 3$ the $X^n$ operator will overtake  the galileon operator at a distance
\be
\frac{\bar r}{r_{\rm v}} \sim \left(\Lambda_{\rm v} r_{\rm v}\right)^\frac{4-4n}{2n-5}\left(\frac{\Lambda_{\rm v}}{\Lambda_\star}\right)^\frac{8n-8}{2n-5}~,
\ee
which is $<1$ for $n\geq 3$.

Conversely, if we are in a situation where the kinetic radius is larger than the Vainshtein radius ({\it i.e.}, a source satisfying~\eqref{largem}) the $X^n$ operator will become important first, and the $\phi'$ profile will be given by~\eqref{xnprofile}; evaluating both $X^n$ and the galileon on this background, we find:
\begin{align}
\Lambda_\star^4 X^n \Big\rvert_{\phi'\sim r^{-2/(2n-1)}} &\sim \Lambda_\star^4\left(\frac{r_\star}{r}\right)^\frac{4n}{2n-1}\\
\frac{1}{\Lambda_{\rm v}}\square\phi(\partial\phi)^2\Big\rvert_{\phi'\sim r^{-2/(2n-1)}} &\sim \Lambda_\star^3 \left(\frac{\Lambda_\star}{\Lambda_{\rm v}}\right)^3 r_\star^{-1} \left(\frac{r_\star}{r}\right)^\frac{2n+5}{2n-1}~.
\end{align}
Here we see that the $X^n$ operator grows more quickly provided that $n \geq 3$. Note that this is exactly the opposite phenomenology from that observed in Section~\ref{sec:intuition}.

\section{Cosmological double screening}
\label{sec:pheno}
After having considered the double screening effects which can occur around isolated point sources, we want to understand whether this phenomenology can be embedded in a consistent cosmological model. Specifically, we want to search for solutions to the model~\eqref{X2plusgal} which are late-time de Sitter attractors; we will not address in detail the approach to these attractors, but leave this for future work. The solutions we construct in this Section are meant primarily as a proof-of-principle that the double screening phenomenology can be made compatible with cosmology. We expect that by generalizing the restricted setup we consider, it will be possible to construct more realistic cosmologies.

One of the primary motivations for considering theories of the galileon type is that they admit cosmological backgrounds which accelerate in the absence of a cosmological constant~\cite{Nicolis:2008in,Chow:2009fm,Silva:2009km,DeFelice:2010pv,Deffayet:2010qz,Barreira:2014jha,Winther:2015pta}. Concretely, if we choose the scale suppressing the galileon operator to be 
\be
\Lambda_{\rm v}^3 \sim H_0^2M_{\rm Pl}\,,
\ee
then the theory admits a solution where the galileon field scales as $\dot\phi \sim H_0M_{\rm Pl}$ and acts as a cosmological constant. Numerically, this fixes the scale $\Lambda_{\rm v} \sim 10^{-40} M_{\rm Pl}$.

Note that the lagrangian~\eqref{X2plusgal} contains an $X^2$ term in addition to the galileon. The simplest phenomenological scenario would be if this operator contributes a subdominant component to the cosmological energy budget, leaving the background evolution to be similar to the well-studied galileon case. However, as we will see, the only way that such a scenario can occur is if $r_\star < r_{\rm v}$ for all collapsed objects,  eliminating the possibility of double screening.
Instead, we will find that the
$X^2$  term must be at least as important  to the cosmological evolution as the galileon for double screening to occur.

In order to gain some intuition for why this is so, consider that the term $(\partial\phi)^4/\Lambda_\star^4$ contributes to the energy density as:
\be
\rho_{X^2}\sim \frac{\dot\phi^4}{\Lambda_\star^4} \sim  \left(\frac{M_{\rm Pl}H_0}{\Lambda_\star}\right)^4.
\ee
If we want this contribution to be subdominant to that of the galileon (which contributes $\rho_{\rm gal.} \sim H_0^2M_{\rm Pl}^2$), we must have:
\be
\Lambda_\star \gg (H_0M_{\rm Pl})^{1/2}.
\label{X2smallerthangal}
\ee
However, this is precisely the condition that the kinetic screening radius be smaller than the Vainshtein radius for an object of mass
\be
M_{\rm Hubble} \sim \rho_{\rm crit.} H_0^{-3},
\ee
which is the total mass enclosed in the present day Hubble volume. Therefore we see that if we satisfy the inequality~\eqref{X2smallerthangal}, we are forced into a situation where the kinetic screening radius is smaller than the Vainshtein radius for all objects in the observable universe. Thus, in order for kinetic screening to be relevant for massive objects, the $X^2$ term must also play a role in the background cosmological evolution, and give a contribution to the background energy density comparable to the galileon. 
\subsection{Background cosmology}
\label{sec:background}
In order to explore the background cosmology---knowing that the $P(X)$-type terms must help drive the background---we generalize slightly the model~\eqref{X2plusgal} to consider an arbitrary $P(X)$ model for the kinetic self-interactions
\be
S = \int\rd^4x\sqrt{-g}\left(\frac{M_{\rm Pl}^2R}{2}+\Lambda_\star^4P(X) - \frac{c_3}{\Lambda_{\rm v}^3}\square\phi(\partial\phi)^2\right),
\label{eq:generalpxcosmoaction}
\ee
where, as before, we define $X \equiv -\frac{1}{2\Lambda_\star^4} (\partial\phi)^2$.
In~\cite{Brax:2014wla,Brax:2014yla}, it was shown that $P(X)$ models are capable of driving accelerated expansion, provided that their small argument expansion is of the form:\footnote{Note that we are essentially introducing a CC by hand into these expressions. In the concrete examples we will consider, $\Lambda$ will be related to the scales $\Lambda_\star$ and $\Lambda_{\rm v}$, so the hope is that hope is that whatever physics is responsible for the presence of the field $\phi$ will also set the CC to be of this form. At the level we are working, this is of course a tuning.}
\be
\Lambda^4_\star P(X) \simeq -\Lambda+c_2 \Lambda_\star^4X - c_4\Lambda_\star^4 X^2+\cdots,
\label{cosmologyaction}
\ee
so we consider the minimal example (we comment on the general case at the end of Section~\ref{sec:pxcosmo}):
\be
\label{eq:mincosmopx}
S = \int\rd^4x\sqrt{-g}\left(\frac{M_{\rm Pl}^2R}{2}-\Lambda-\frac{c_2}{2}(\partial\phi)^2 - \frac{c_3}{\Lambda_{\rm v}^3}\square\phi(\partial\phi)^2 - \frac{c_4}{4\Lambda_\star^4}(\partial\phi)^4\right).
\ee
The desire to have $r_\star > r_{\rm v}$ for the most massive clusters also places a bound on the ratio between the scales in the theory. A basic requirement is that $r_\star > r_{\rm v}$ for a source of the mass enclosed in a Hubble volume, which leads to the following bound on the ratio $\Lambda_\star/\Lambda_{\rm v}$:
\be
\frac{\Lambda_\star}{\Lambda_{\rm v}} < \left(\frac{g M_{\rm Pl}}{H_0}\right)^{1/6}\,.
\label{havingrsgrv}
\ee
For the canonical situation with $H_0\sim 10^{-60}M_{\rm Pl}$ and an ${\cal O}(1)$ coupling to matter, this corresponds to the ratio of scales being smaller than $10^{10}$. To begin, we will search for solutions where $\Lambda = 0$, but will see that these solutions do not allow for double screening of collapsed objects, though they do admit a cosmological analogue of double screening.

The theory~\eqref{eq:mincosmopx} has an exact shift symmetry, so the equation of motion for the scalar field is equivalent to conservation of the current associated to this symmetry, $\nabla^\mu J_\mu = 0$, with~\cite{Deffayet:2010qz}
\be
J_\mu = \left(c_2 -2 c_4X+\frac{2c_3}{\Lambda_{\rm v}^3}\square\phi\right)\partial_\mu\phi+\frac{2c_3\Lambda_\star^4}{\Lambda_{\rm v}^3}\partial_\mu X.
\ee
Additionally, the stress tensor following from the action~\eqref{cosmologyaction} is 
\begin{align}
\nonumber
T_{\mu\nu} =&-\Lambda g_{\mu\nu}+c_2\partial_\mu\phi\partial_\nu\phi +c_2\Lambda_\star^4Xg_{\mu\nu}\\
&+\frac{2c_3\Lambda_\star^4}{\Lambda_{\rm v}^3}(\partial_\mu\phi\partial_\nu X+\partial_\nu\phi\partial_\mu X)- g_{\mu\nu}\frac{2 c_3\Lambda_\star^4}{\Lambda_{\rm v}^3}\partial^\alpha\phi\partial_\alpha X+\frac{2c_3}{\Lambda_{\rm v}^3}\square\phi\partial_\mu\phi\partial_\nu\phi\\\nonumber
&- 2c_4X\partial_\mu\phi\partial_\nu\phi-c_4\Lambda_\star^4X^2g_{\mu\nu}.
\end{align}
We are interested in cosmological solutions to the equations of motion. We therefore make an FLRW ansatz for the metric and specialize to homogeneous profiles for the scalar field $(\phi = \phi(t))$. We further note that the scalar equation of motion admits a first integral, so that the combined equations of motion for the scalar and the metric may be cast as
\begin{align}
\label{friedmanneq1}
3M_{\rm Pl}^2H^2 &= \frac{c_2}{2}\dot\phi^2- \frac{6H c_3}{\Lambda_{\rm v}^3}\dot\phi^3 - \frac{3c_4}{4}\frac{\dot\phi^4}{\Lambda_\star^4}+\Lambda \\\label{friedmanneq2}
3M_{\rm Pl}^2H^2+2M_{\rm Pl}^2\dot H &= -\frac{c_2}{2}\dot\phi^2 -\frac{2c_3}{3\Lambda_{\rm v}^3}\frac{\rd}{\rd t}\dot\phi^3+ \frac{c_4}{4\Lambda_\star^4}{\dot\phi^4}+\Lambda\\
c_2\dot\phi -\frac{6c_3 H}{\Lambda_{\rm v}^3}\dot\phi^2-\frac{c_4}{\Lambda_\star^4}\dot\phi^3 &= \frac{C}{a^3},
\label{kgeq}
\end{align}
where equations~\eqref{friedmanneq1} and~\eqref{friedmanneq2} are the Friedmann equations and~\eqref{kgeq} is the integrated scalar equation of motion, with $C$ an integration constant. Only 2 of these equations are independent, the third may be derived from the others. In much of what follows, it will be more convenient to work in terms of the variable $X = \dot\phi^2/(2\Lambda_\star^4)$ rather than $\dot\phi$; we just must ensure that $X > 0$ so that $\dot\phi$ is real. 

We search for exact de Sitter solutions, satisfying $X = {\rm const.}$ and $H = {\rm const.}$; we assume that at late times, the system approaches an attractor where $C\to 0$. This is reasonable, because as the universe expands, $C/a^3$ will be driven to zero. Using the same method as~\cite{Kobayashi:2010cm}, we can manipulate the background equations to obtain an equation solely for $X$. To do this, we first subtract~\eqref{kgeq} from~\eqref{friedmanneq1} to obtain:
\be
H^2 = \frac{\Lambda_\star^4}{3M_{\rm Pl}^2}\left(c_4X^2 - c_2 X+ \ell\right),
\label{eq:hubble1}
\ee
where we have defined $\Lambda = \ell \Lambda_\star^4$. Similarly, we can solve~\eqref{kgeq} for $H$ to obtain:
\be
H^2 = \frac{\Lambda_{\rm v}^6}{72 c_3^2 \Lambda_\star^4 X}\left(c_2-2c_4X\right)^2.
\ee
In order to obtain an equation solely for $X$, we can equate these two expressions to yield
\be
\frac{\left(c_2-2c_4 X\right)^2}{X\left(c_4X^2-c_2X+\ell\right)} = \frac{24 c_3^2\Lambda_\star^8}{\Lambda_{\rm v}^6 M_{\rm Pl}^2}.
\label{eq:kgX}
\ee
We therefore want to solve this equation for $X$, and then plug the value back into~\eqref{eq:hubble1} to obtain an expression for $H$. The solutions are not totally unconstrained however, we demand that they be perturbatively stable and have solutions which screen around collapsed objects. In the next Section, we derive the constraints that these considerations place upon the theory before explicitly solving the background equations.

\subsection{Perturbations and coupling to matter}
\label{sec:perturbations}
In order to evaluate the viability of solutions to the background equations derived in~\eqref{sec:background}, a zeroth order check is that they are perturbatively stable. Further, we would like fluctuations about these solutions to be screened in the presence of massive sources. In this Section, we give the action governing linear perturbations about cosmological solutions in the theory~\eqref{cosmologyaction} and the action governing perturbations in the quasi-static regime so that we can investigate constraints on the theory placed by the above considerations.

\subsubsection{Perturbative stability}
We begin by considering perturbations about the cosmological solutions. In order to do this, we perturb both the scalar $\phi = \bar\phi+\delta\phi$ and the metric $g_{\mu\nu} = \bar g_{\mu\nu} +h_{\mu\nu}$ and we work in ADM variables for the metric:
\be
\rd s^2 = -N^2 \rd t^2 + h_{ij}\left(\rd x^i+N^i\rd t\right)\left(\rd x^j+N^j\rd t\right),
\ee
and choose the gauge: $ \delta\phi = 0; h_{ij}=a^2 e^{2\zeta}(e^\gamma)_{ij}; \gamma^i_i = \partial_i\gamma^{ij} = 0$.
Substituting this parameterization of the metric into the action, solving for the lapse and shift at first order, and then substituting back into the action yields the quadratic action for $\zeta$~\cite{Deffayet:2010qz,Kobayashi:2010cm,Burrage:2011hd},
\be
S = \int\rd^4x~a^3{\cal Z}^2\left(\dot\zeta^2
- \frac{c_s^2}{a^2}(\vec\nabla\zeta)^2\right),
\label{pertaction}
\ee
where the scale factor is that of de Sitter, and the coefficients are given by
\begin{align}
\label{eq:z2}
{\cal Z}^2 =&~\Lambda_\star^4X\left(H+\frac{2c_3\Lambda_\star^4}{\Lambda_{\rm v}^3 M_{\rm Pl}^2}\dot\phi X\right)^{-2}\left(-c_2-2c_4X+\frac{X(c_2-2c_4X)^2}{(-c_2X+c_4X^2+\ell)}\right)\\
\label{eq:cs2}
c_s^2 =& \frac{\left(c_2-2c_4 X\right)\left(c_4 X^2 - \ell\right)}{6c_2^2 X+6 c_4 X\left(c_4 X^2 -\ell\right)-3 c_2\left(3 c_4 X^2+\ell\right)},
\end{align}
where we have used the background equations of motion to simplify these expressions.
We can use these expressions to investigate the stability of cosmological solutions. In particular, in order to avoid ghost instabilities, we demand ${\cal Z}^2 > 0$; to avoid gradient instabilities, we demand $c_s^2 > 0$. 

\subsubsection{Induced coupling to matter}
One of the more interesting aspects of the galileon is that its higher-derivative structure naturally induces couplings between the field $\phi$ and matter sources, essentially through $\phi$ coupling to curvature. This occurs even in the absence of an explicit coupling between $\phi$ and matter at the level of the action. In this section, we would like to estimate the magnitude of this coupling on the cosmological backgrounds of interest.

The equation of motion for the scalar $\phi$ can be written as
\be
\square\phi+\frac{2c_3}{\Lambda_{\rm v}^3}\left((\square\phi)^2-(\nabla_\mu\nabla_\nu\phi)^2 - R_{\mu\nu}\nabla^\mu\phi\nabla^\nu\phi\right) - 2c_4\nabla_\mu\left(X\nabla^\mu\phi\right) = 0\,.
\ee
We perturb both the scalar $\phi = \bar\phi(t)+\varphi$ and the metric, working in Newtonian gauge where the line element takes the form:
\be
\rd s^2 = -(1+2\Psi)\rd t^2+a^2(1-2\Phi)\rd\vec x^2.
\ee
In order to understand the physics around isolated objects on the cosmological background, it suffices to work in the quasi-static approximation, where we neglect terms of order $\dot\varphi$ as well as terms proportional to the gravitational potentials, $\Phi, \Psi$ or their first derivatives $\nabla\Phi,\dot\Phi, \nabla\Psi,\dot\Psi$.\footnote{We assume that nothing subtle happens at scales intermediate between cosmological and quasi-static scales, but it should in principle be checked that structures can evolve in such a way that they approach the quasi-static regime in a stable way. See~\cite{Babichev:2012re} for an investigation along these lines in the cubic galileon.} Further, we neglect terms which have only linear gradients of $\varphi$ relative to those with second gradients---$\nabla^2\vp/\Lambda_\star^2 \gg \nabla\vp/\Lambda_\star$.
In this approximation, the equation of motion for $\varphi$ becomes~\cite{Brax:2014yla,Winther:2015pta}
\be
\beta\nabla^2\vp+ \frac{2c_3}{ \Lambda_{\rm v}^3 a^2}\left((\nabla^2\vp)^2 - (\nabla_i\nabla_j\vp)^2\right)+\frac{c_4}{\Lambda_\star^4 a^2}\nabla_i\left(\nabla^i\vp(\vec\nabla\vp)^2\right) = \frac{2a^2c_3\Lambda_\star^4{\bar X}}{\Lambda_{\rm v}^3M_{\rm Pl}^2}\delta\rho,
\label{eq:qstatic}
\ee
with
\be
\label{eq:beta}
\beta = \frac{(c_2-2c_4 X)(c_4 X^2-\ell)}{3(-c_2 X+c_4 X^2 +\ell)}.
\ee
We see that the strong coupling scales are re-dressed by powers of $\beta$, while the effective coupling to matter is given by
\be
g_{\rm eff}\sim\frac{\Lambda_\star^4{\bar X}}{\beta\Lambda_{\rm v}^3M_{\rm Pl}}.
\ee
One important thing to notice is that $\beta \propto c_s^2$, where the proportionality factor is positive if ${\cal Z}^2 > 0$, which we will demand. We therefore see that on viable cosmological solutions, $\beta > 0$, which constrains the sign that we can have for $c_4$ while still allowing screening.\footnote{There is no similar constraint on the sign of $c_3$ because the sign of the coupling to matter is also controlled by $c_3$.} Comparing~\eqref{eq:qstatic} to~\eqref{eq:screeningpointsource}, we see that $c_4 = 1$ is necessary to have screening around collapsed objects for very massive sources where kinetic screening dominates. We are now ready to construct explicit cosmological solutions and investigate their properties.

\subsection{Cosmological solutions}
\label{sec:pxcosmo}
We now want to solve the equations~\eqref{eq:hubble1} and~\eqref{eq:kgX} and check that the resulting cosmologies are stable and allow for screening around compact objects. We first consider the situation without an explicit cosmological constant.

\paragraph{No cosmological constant:}

We look for exact de Sitter solutions in the absence of a cosmological constant by setting $\ell = 0$. As was mentioned above, in order to have $r_\star > r_{\rm v}$ for the largest objects, we would like for the cosmological background to be driven by the energy density of the $X^2$ term.

Looking at equation~\eqref{eq:kgX}, we see that so long as $\varepsilon \equiv \Lambda_\star^4/(\Lambda_{\rm v}^3 M_{\rm Pl})\ll 1$,
we can solve for $X$ by perturbing in this small quantity. Note that by squaring this quantity, we see that this requirement is equivalent to
\be
\left(\frac{\Lambda_\star}{\Lambda_{\rm v}}\right)^6\left(\frac{\Lambda_\star}{M_{\rm Pl}}\right)^2 \ll1,
\label{perturbativeineq}
\ee
but if we imagine that the $X^2$ term is responsible for driving the background, we expect that $M_{\rm Pl}^2H_0^2 \sim \Lambda_\star^4$, which implies that we can rewrite the inequality~\eqref{perturbativeineq} as
\be
\left(\frac{\Lambda_\star}{\Lambda_{\rm v}}\right)^6\left(\frac{\Lambda_\star}{M_{\rm Pl}}\right)^2 \ll \frac{M_{\rm Pl}}{H_0},
\ee
which is essentially the requirement to have $r_\star > r_{\rm v}$ inside the horizon~\eqref{havingrsgrv}. Therefore, we expect that $\varepsilon$ will be small whenever we have double screening within the horizon, thus this is the relevant limit to work in.
We therefore expand $X = X_0+\varepsilon X_1+\cdots$; 
 at leading order, we find that $X_0 = c_2/(2c_4)$. Recalling that we must have $X \geq 0$ in order for $\dot\phi$ to be real, we see that $c_2$ and $c_4$ must have the same sign. At the next order, we find $X_1^2 = -(3c_2^3 c_3^2\Lambda_\star^8)/(4c_4^4\Lambda_{\rm v}^6M_{\rm Pl}^2)$. In order for $X_1$ itself to be real, we must therefore have $c_2 = -1$, which implies that $c_4 = -1$ as well, so that the solution for $X$ is given by
\be
X = \frac{1}{2} \pm c_3\sqrt\frac{3}{4}\frac{\Lambda_\star^4}{\Lambda_{\rm v}^3M_{\rm Pl}}+\cdots,
\label{noccXsoln}
\ee
this can then be substituted into~\eqref{eq:hubble1} to find that the Hubble parameter is given by
\be
H \simeq \frac{1}{\sqrt{12}}\left(\frac{\Lambda_\star}{M_{\rm Pl}}\right)\Lambda_\star.
\ee
We can insert these expressions into~\eqref{eq:z2},~\eqref{eq:cs2} and~\eqref{eq:beta} to obtain\footnote{We are free to choose either branch for $X_1$ in~\eqref{noccXsoln}, but the sign of $c_3$ must then be chosen to make $c_s^2 >0$, so these two signs cannot be chosen independently.}
\begin{align}
{\cal Z}^2 &\simeq 12M_{\rm Pl}^2+\cdots > 0\\
c_s^2 &\simeq \mp\frac{c_3}{2\sqrt 3} \frac{\Lambda_\star^4}{\Lambda_{\rm v}^3 M_{\rm Pl}}+\cdots > 0\\
\beta &\simeq \mp\frac{c_3}{\sqrt 3} \frac{\Lambda_\star^4}{\Lambda_{\rm v}^3 M_{\rm Pl}}+\cdots > 0 \implies g_{\rm eff} \sim {\cal O}(1).
\end{align}
We see that for there are indeed stable self-accelerated solutions in the absence of a CC (indeed, these are precisely the solutions used for inflation in~\cite{Kobayashi:2010cm}). On these backgrounds, the $X^2$ term is the dominant source of stress energy, with the galileon playing a negligible role for both the background and on linear scales. However, these solutions will {\it not} exhibit kinetic screening around isolated point sources, because $c_4 = -1$.

This fact does not immediately rule these solutions out. For modest values of $\varepsilon$, the kinetic screening radius will only be larger than the Vainshtein radius for objects which are a significant fraction of the total mass in a Hubble volume. There are no collapsed objects of such large masses, so the theory will still be phenomenologically safe; all collapsed objects will be Vainshtein screened. An extreme limit of this is when $\varepsilon\sim1$---which happens for $\Lambda_\star/\Lambda_{\rm v}\,\gsim\, 10^{10}$---where $r_\star$ will only be larger than $r_{\rm v}$ for objects as heavy as all the mass enclosed in our Hubble volume. In this regime, perturbation theory in $\varepsilon$ obviously breaks down and a solution must be found by other means.

The situation where $X^2$ drives the background but objects are Vainshtein screened is more interesting than it might na\"ively appear; it is something of a cosmological analogue of double screening, where linear perturbations in cosmology are governed by the $X^2$ term while the dynamics of collapsed objects is controlled by the galileon. This possibility has not been extensively studied before in the literature, this provides a way for $P(X)$ theories to self-accelerate using the ghost condensate~\cite{ArkaniHamed:2003uy} mechanism, but still satisfy local tests of gravity, even in the presence of a coupling to matter. Note also that this illustrates the point that inferring parameters of dark energy by observing perturbations on linear scales may cause erroneous inferences about the physics governing small scale effects in the laboratory or Solar System (and vice versa) because the same operators do not necessarily dominate the physics on all scales.

Therefore, we see that this is a phenomenologically viable possibility where the statistics of linear fluctuations is controlled by a different operator than that which controls the dynamics of bound objects.
However, we would like to see if it is possible to arrange a situation where the physics of {\it collapsed} objects is different, depending on their mass.
For the restricted theory~\eqref{X2plusgal}, this appears to require introducing a bare cosmological constant.

\paragraph{With a cosmological constant:}

We now want to solve the same set of equations, but with $\ell \neq 0$. In order to have kinetic screening, we must have $c_4 = 1$. Our first inclination is to try to set $c_2 = 1$ as well. However, it is not possible to satisfy $c_s^2 > 0, {\cal Z}^2 > 0$ simultaneously for any choice of $\ell$ for $c_2 = 1$. We therefore consider the situation with $c_2 = -1$.

In this case, we want to perform a similar perturbative expansion as in the case without a CC, but we will count $\ell$ as starting at order $\varepsilon^{-2}$. That is, we expand $\ell = \varepsilon^{-2}\ell_{-2}+\varepsilon^{-1}\ell_{-1} + \cdots$. The intuition for working in the large $\ell$ limit is that both $c_s^2$ and ${\cal Z}^2$ greatly simplify and it is straightforward to see that they are positive with the sign choices we have made. By solving~\eqref{eq:kgX} perturbatively, we find to leading order in $\varepsilon$:
\begin{align}
X &\simeq X_0+\cdots\\
\ell &\simeq \frac{(1+2X_0)^2\Lambda_{\rm v}^6 M_{\rm Pl}^2}{24 c_3^2X_0\Lambda_\star^8}\\
H &\simeq \pm \frac{(1+2X_0)}{6c_3\sqrt{2X_0}}\left(\frac{\Lambda_{\rm v}}{\Lambda_\star}\right)^2\Lambda_{\rm v}+\cdots,
\end{align}
so we see that these solutions require a cosmological constant of order $\Lambda \sim \Lambda_{\rm v}^6M_{\rm Pl}^2/\Lambda_\star^4$, which is parametrically larger than $\Lambda_\star^4$. We are free to set the ratio $\Lambda_\star/\Lambda_{\rm v}$ to be small enough to allow small mass objects to be Vainshtein screened while having heavy objects be kinetic screened. On these solutions, the galileon and  $X^2$  
 terms have comparable contributions to the background energy density, both being subdominant to the CC; similarly both terms contribute to linear perturbations.

In order to check the perturbative stability of these solutions, we compute
\begin{align}
{\cal Z}^2 &\simeq \frac{72c_3^2(1-2X_0)X_0^2}{(1+2X_0)}\left(\frac{\Lambda_\star}{\Lambda_{\rm v}}\right)^6\Lambda_\star^2+\cdots > 0\\
c_s^2 &\simeq \frac{1+2X_0}{3-6X_0}+\cdots > 0\\
\beta &\simeq \frac{1}{3}(1+2X_0)+\cdots > 0 \implies g_{\rm eff} \sim \frac{3X_0^2}{(1+2X_0)}\frac{3\Lambda_\star^4}{\Lambda_{\rm v}^3M_{\rm Pl}}.
\end{align}
Stability of these solutions requires $X_0 < 1/2$. Note also on these solutions that $g_{\rm eff} \ll 1$, this means that if we want to have a large effect from the presence of the $\phi$ field ({\it i.e.}, for it to mediate a gravitational strength fifth force) we should introduce an explicit coupling between $\phi$ and matter, in addition to the coupling induced by the galileon term. Indeed, the coupling $g_{\rm eff}$ is so small, that it causes~\eqref{havingrsgrv} to not be satisfied, so we must introduce an explicit coupling to matter in order to have double screening. 

\begin{figure}
\centering
\includegraphics[width=3.2in]{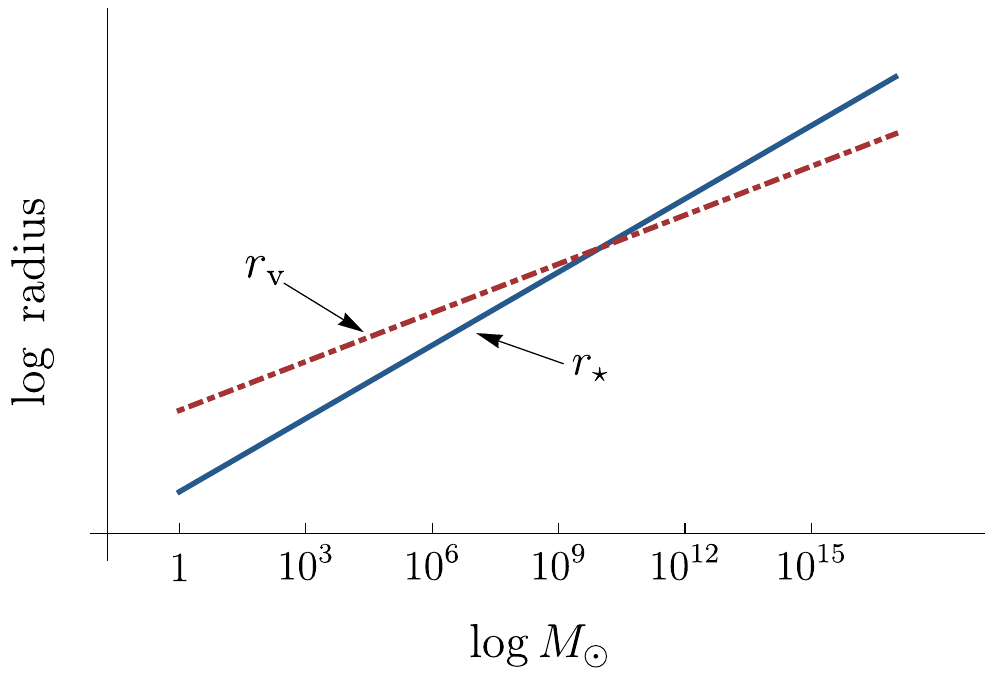}
\caption{\label{fig:rvrsvM}\small Log-log plot of kinetic screening radius (blue solid) and Vainshtein radius (red dot dashed) as a function of the mass of an isolated source for the double screening cosmology in the presence of a CC. Here we have chosen the fiducial parameters $\Lambda_\star = 10^{-34} M_{\rm Pl}$, $\Lambda_{\rm v} = 10^{-42} M_{\rm Pl}$ corresponding to a crossover mass of $M \sim 10^{10}M_\odot$ and a Hubble constant of $H \sim 10^{-59} M_{\rm Pl}$. We see in the plot that large mass objects are kinetic screened, while low mass objects are Vainshtein screened, confirming that we can embed the double screening phenomenology in a cosmological scenario. Note that in this case the induced coupling to matter from the galileon is very suppressed, so we have assumed that an explicit ${\cal O}(1)$ coupling to matter is present.}
\end{figure}

In Figure~\ref{fig:rvrsvM} we plot the Vainshtein radius and kinetic screening radius as a function of mass for a fiducial choice of parameters in this family of solutions, confirming that it is possible to have different screening behavior for different bound objects.

\paragraph{Cosmology with more general $P(X)$:} In this Section, we have focused on the cosmology of the specific model~\eqref{eq:mincosmopx}, here we comment on generalizations to include other operators of the $P(X)$ type, as in Section~\ref{sec:px}. Concretely, we consider the cosmology of the theory~\eqref{eq:generalpxcosmoaction}. For this general action, the equations of motion are given by~\cite{Deffayet:2010qz}
\begin{align}
&\dot\phi\left(P_{,X}-6Hc_3\dot\phi\Lambda_{\rm v}^{-3}\right)=\frac{C}{a^3}\\
&3M_{\rm Pl}^2H^2 = \Lambda_\star^4\left(2XP_{,X}-P\right)-\frac{6Hc_3}{\Lambda_{\rm v}^3}\dot\phi^3\,.
\end{align}
These equations at late times also admit an attractor, $C\to 0$. For particular choices of $P(X)$, these equations should admit de Sitter solutions on these attractors.

A particularly interesting class of solutions consists of the galileon operator driving the background expansion, while having the $P(X)$ terms be a subdominant contribution to the energy density. 
This set-up allows for a different type of double screening. 
For concreteness, consider the situation with the galileon plus an operator ${\cal L} \sim X^3$. 
This operator grows more quickly than the galileon as we approach massive objects, and when it is the dominant operator, the force due to $\phi$ scales as $\phi'\sim r^{-2/5}$. 
We imagine that the galileon drives the background while the $X^3$ operator is a negligible correction (and hence $r_{\rm v} > r_\star$ everywhere). 
In this case, far from sources initially things will be Vainshtein screened, but now the $X^3$ operator grows faster than the galileon, so screening will switch over to kinetic screening at smaller radii. 
This is a different type of double screening, where all objects behave in the same way, but there are two different screening regimes around all objects. 
This might be of interest for laboratory investigations.

It should also be possible in this case to find solutions where the $P(X)$ term is responsible for driving the background, which would exhibit a similar type of double screening phenomenon discussed above. However, in this case, large mass objects would be kinetic screened all the way to $r = 0$, while small mass objects will have two regimes, being first Vainshtein screened and then kinetic screened closer to the source.
In this case, the phenomenology is essentially exactly opposite that of the $X^2$ plus galileon example we have considered for the majority of the text.

For completeness, we note that perturbations about cosmological solutions will be governed by an action similar to~\eqref{pertaction}, but now with~\cite{Deffayet:2010qz,Kobayashi:2010cm,Burrage:2011hd}
\begin{align}
{\cal Z}^2 &=~\Lambda_\star^4X\left(H+\frac{2c_3\Lambda_\star^4}{\Lambda_{\rm v}^3 M_{\rm Pl}^2}\dot\phi X\right)^{-2}\left(-P_{,X}+2XP_{,XX}-\frac{XP_{,X}^2}{P}\right)\\
c_s^2 &= \frac{P_{,X}-XP_{,X}^2/P}{3\left(P_{,X}-2XP_{,XX}+XP_{,X}^2/P\right)}.
\end{align}
Note that we have eliminated most of the dependence on the galileon operator by utilizing the background equations of motion, which obfuscates taking the pure $P(X)$ limit.
By allowing for a general functional form for $P$---and possibly higher-order galileon operators---it should be able to find phenomenologically interesting solutions which are at the same time perturbatively stable.

\section{Radiative stability}
\label{sec:rad}
Thus far, we have focused on the phenomenology of double screening. However, in order to be confident that the behaviors we have uncovered are robust, we should ensure that the starting point---the choice of action for the scalar field---is quantum-mechanically stable. If quantum corrections are large, we are not justified in truncating the action in the manner we have chosen and neglecting higher-order operators.

As a fiducial model, we investigate the radiative stability of the theory 
\be
{\cal L}=-\frac{1}{2}(\partial\phi)^2 -\frac{1}{\Lambda_{\rm v}^3}\square\phi(\partial\phi)^2 - \frac{1}{\Lambda_\star^4}(\partial\phi)^4,
\label{pXgal}
\ee
and then comment on the generalization to $P(X)$ theories, which is straightforward. We investigate the quantum corrections to this action in two ways: by first presenting a heuristic power-counting argument, and then by computing the 1-loop Coleman--Weinberg potential~\cite{Coleman:1973jx} using heat kernel techniques~\cite{Barvinsky:1985an,Barvinsky:1990up}. 

As has been recently argued in~\cite{Pirtskhalava:2015nla}, this model fits into a class of quantum-mechanically well-behaved theories, which are distinguished amongst theories of the Horndeski type~\cite{Horndeski:1974wa}. The radiative stability of the model can be traced to the fact that in the limit that $\Lambda_\star \gg \Lambda_{\rm v}$, the theory displays an enhanced symmetry---that of the galileon $\delta\phi = x^\mu$. For this reason~\cite{Pirtskhalava:2015nla} dubbed this ``weakly-broken galileon symmetry.'' It is known that in the theory of the galileon, the 1PI effective action for the galileon 
contains only additional operators of the form~\cite{Nicolis:2004qq,Luty:2003vm,Hinterbichler:2010xn}
\be
\Delta{\cal L} \sim \frac{\partial^m(\partial^2\phi)^n}{\Lambda_{\rm v}^{m+3n-4}}~,
\ee
so we expect that operators of the $P(X)$ form will be suppressed by some ratio of $\Lambda_{\rm v}/\Lambda_\star$, as they must vanish in the limit where $\Lambda_{\rm v}/\Lambda_\star\rightarrow 0$. Note that in this picture, the radiative stability of the model can be traced back to the large hierarchy between the scales $\Lambda_\star$ and $\Lambda_{\rm v}$.

\subsection{Power counting argument}

We would like to show that radiative corrections to~\eqref{pXgal} of the $P(X)$ type---that is, with one derivative per field---come suppressed not by the scale $\Lambda_\star$, but by a parametrically higher scale (see also~\cite{Pirtskhalava:2015nla}). There is a power-counting argument that shows this---essentially the same as that of~\cite{Creminelli:2012my}; the essence of the argument is that galileon vertices must always connect internal lines in a given diagram and therefore contribute positive powers of $\Lambda_{\rm v}$ to the overall scaling.

In a generic loop diagram, the elements can be separated into internal lines (and the vertices which connect only to internal lines) and external lines (and vertices which connect to them).
A given term in the 1PI effective action will then take the following form:
\be
{\cal O}_{m,n,q} = \left(\frac{\Lambda_{\rm v}}{\Lambda_\star}\right)^q\frac{\partial^m(\partial\phi)^n}{\Lambda_\star^{2n+m-4}}\,.
\ee
What we want to argue is that $q > 0$ for all terms with $m=0$. First, note that if galileon vertices connect to external lines, the external lines would have at least 2 derivatives per field (this follows from the galileon non-renormalization theorem, see {\it e.g}.,~\cite{Hinterbichler:2010xn}). Therefore the galileon vertices connect only to internal lines.

For internal lines/vertices,  all of the momenta are integrated over. If we take the strong coupling scale of the theory, $\Lambda_{\rm v}$, to be the {\it true} cutoff 
and if we employ a hard-momentum cut-off regularization, then at most the momenta running in the loops will be of order $\Lv$.  The internal elements will then contribute positive powers of $\Lambda_{\rm v}$---as they are UV divergent---and possibly negative powers of $\Lambda_\star$ coming from $X^2$ vertices, so that overall they have $q >0$. 

Next, consider the vertices connected to external lines. These must come from the $X^2$ interaction, and are therefore suppressed by powers of $\Lambda_\star$. These elements will have no powers of $\Lambda_{\rm v}$ and therefore have $q=0$. Thus, the total $q$ for a generic graph at $m=0$ will have $q>0$, as expected. This implies that terms in the quantum effective action of the $P(X)$ type come suppressed by a scale parametrically higher than $\Lambda_\star$.

The previous argument reduces roughly to the fact that the only way we could get negative powers of $\Lambda_{\rm v}$ from a graph would be to have a galileon vertex connect to an external line; but this type of graph will always have at least 2 derivatives per external leg. Though we have focused here on the theory~\eqref{pXgal}, the above arguments extend straightforwardly to the general case~\eqref{eq:galileonandgeneralpx}
and even beyond~\cite{Pirtskhalava:2015nla}. 
\subsection{1-loop effective action}
\label{subsec:BV}
The previous argument for radiative stability is satisfactory at energies below or about the scale $\Lambda_{\rm v}$, which we have interpreted as a true cutoff of the theory. However, we would like to trust the theory beyond this scale; indeed, we would like to be able to consider situations where the background satisfies both
\be
\left|\frac{\partial^2\bar{\phi}}{\Lambda_{\rm v}^3} \right| \gg 1~~~~~~~~{\rm and}~~~~~~~~\left|\frac{(\partial\bar{\phi})^2}{\Lambda_\star^4}\right| \gg 1,
\label{largebkgd}
\ee
in various regimes; that is, situations where screening can occur due to either the galileon or the kinetic screening term.

It is indeed possible to do this in a robust way, but we must be slightly less ambitious than in the previous section. There, the theory is quantum-mechanically stable even accounting for power law divergences in the $\phi$ sector. These divergences are somewhat model-dependent in that they strongly depend on the nature of the UV completion of the low energy theory in which they arise. Though they often accurately forecast the influence of heavy states that have been integrated out, taking the power law divergences seriously sometimes overestimates the magnitude of quantum corrections~\cite{Burgess:1992gx}. Therefore, we will instead make the more modest demand that quantum corrections associated to~{\it logarithmic} divergences are under control. This does make optimistic assumptions about the nature of any putative UV completion of these theories, but it seems to us to at least be a necessary---if not sufficient---condition for the theory to be well-behaved.

It should not be surprising that the theory is under quantum control in this sense, as the backgrounds of interest are quite similar to the screening solutions of either a $P(X)$ model or the galileon, combined in a novel way. Despite this, only one of these types of operators dominates in any given regime, and it is known that these theories are individually well-behaved with respect to logarithmic divergences in these regimes.

In order to see this, we compute the 1PI effective action, $\Gamma[\phi]$, for $\phi$ in the theory~\eqref{eq:galileonandgeneralpx} via the background field method. We split the field into a background and perturbation (not necessarily small) as $\phi = \bar\phi+\varphi$ and integrate out $\vp$ at 1-loop order, which requires only keeping terms up to quadratic order in $\varphi$ in the path integral:
\be
e^{i\Gamma[\bar\phi]} = \int_{\rm 1PI}{\cal D}\varphi ~\exp\left(-i\int\rd^4x\frac{1}{2}Z^{\mu\nu}(\bar\phi)\partial_\mu\varphi\partial_\nu\varphi\right)\,,
\ee
where the kinetic matrix, $Z^{\mu\nu}$, is given by
\be
Z^{\mu\nu} = \left(\frac{4}{\Lambda_{\rm v}^3}\square\bar\phi+2\bar P'\right)\eta^{\mu\nu}-\frac{4}{\Lambda_{\rm v}^3}\partial^\mu\partial^\nu\bar\phi-\frac{4}{\Lambda_\star^4}\bar P'' \partial^\mu\bar\phi\partial^\nu\bar\phi~,
\ee
and barred quantities are to be understood as being evaluated on the background $\bar{\phi}$.

The 1-loop computation we want to do can be cast as a geometric expansion using heat kernel techniques~\cite{Parker,Tetradis:1993ts,Berges:2000ew,Avramidi:2001ns,Codello:2012dx,Brouzakis:2013lla}.  The part of the 1PI effective action associated to logarithmic divergences  arranges itself into the following curvature invariants~\cite{Barvinsky:1985an,Barvinsky:1990up,deRham:2014wfa}
\be
\Gamma_{{\rm 1-loop}}[\bar\phi] \sim \int\rd^4 x\sqrt{G_{\rm eff}}\left(R^2 +2R_{\mu\nu}^2\right)~,
\label{eq:1looplog}
\ee
defined in terms of the effective metric
\be
G^{\rm eff}_{\mu\nu}\equiv \dfrac{1}{\sqrt{Z}}\ Z_{\mu\nu}\ .
\ee
This contribution is independent of regularization procedure. Furthermore, being recast as a covariant contribution to the EFT lagrangian, it is appropriate for backgrounds that break Lorentz invariance. Other contributions will either depend on the regularization scheme used and the value of the cut-off of the theory (about which we have no knowledge), or they will be finite, harmless contributions.
In terms of the kinetic matrix, $Z$, the curvature terms scale as $R\sim \partial^2\log Z$~\cite{Nicolis:2004qq}. 
The conditions for the theory to be radiatively stable are then\footnote{At this point we are ignoring the matter sector the theory couples to in order for the scalar field $\phi$ to acquire a large background value, $\bar{\phi}$. Loop corrections arising from dynamical matter coupled to scalar field theories are typically large and unsuppressed. At this level we content ourselves in showing the consistency of the theory coupled to non-dynamical matter sources which do not run in the loops.}
\be
\frac{\partial^2 Z}{Z}~~{\rm and}~~\left(\frac{\partial Z}{Z}\right)^2 \ll \Lambda_\star^4 P(X) -\frac{1}{\Lambda_{\rm v}^3}\square\phi(\partial\phi)^2~.
\ee
If we have a classical background where~\eqref{largebkgd} holds, the theory will be under control so long as (in somewhat schematic notation)
\be
\frac{\partial }{\Lambda_\star} \ll 1~~~~{\rm and}~~~~~\frac{\partial}{\Lambda_{\rm v}}\ll 1\,,~~~~{\rm in~the~regime}~~~~\bar{X} \gtrsim 1~~~~{\rm and}~~~~~\frac{\Box\bar{\phi}}{\Lambda_{\rm v}^3}\gtrsim 1\,.
\ee
This can be checked
explicitly upon replacing the background screening solution for $\bar{\phi}$. On the screening solutions of interest, these parameters scale as
\be
\frac{\partial }{\Lambda_\star} \sim \frac{1}{r\Lambda_\star}~~~~{\rm and}~~~~~\frac{\partial}{\Lambda_{\rm v}}\sim \frac{1}{r\Lambda_{\rm v}}.
\ee
Neither of these becomes ${\cal O}(1)$ until we reach $r\sim \Lambda_{\rm v}^{-1}$, which is parametrically smaller than both $r_\star$ and $r_{\rm v}$.

To be more precise, in the regime when $\bar{X} \gtrsim 1$, the EFT lagrangian will receive contributions which will scale at worst as $(\phi''(r))^m/\Lstar^{3m-4}$, where $m$ is a positive integer. Since these contributions only become important at the higher energy scale $\Lstar$, they provide very small corrections to the Galileon operators. 

In this 
%note 
analysis we have focused on the region of parameter space where $\Lambda_\star \gg \Lambda_{\rm v}$, owing to its nice quantum behavior which is protected by approximate galileon symmetry. However, it may be that there are other interesting regions of parameter space which are also stable in this sense. One promising situation is the opposite hierarchy $\Lambda_{\rm v} \gg \Lambda_\star$, as it is known that pure $P(X)$ is by itself stable with respect to logarithmic divergences~\cite{deRham:2014wfa}. However, generic kinetic interactions in that case will generate operators 
which will give contributions to the equations of motion of the same order as the Galileon operators. 

\paragraph{Higher loops and UV physics:} 

To this point, we have ignored the effects of heavy UV physics (by focusing on log divergences in the effective action). It is, of course, desirable to move beyond this approximation---since the UV theory is unknown, the only handle we have on these effects is through power-law divergences, which serve as a proxy for the high-energy physics.  We therefore would like to estimate these effects in the theory~\eqref{eq:galileonandgeneralpx}. Alternatively, one could explicitly integrate out a heavy matter field and focus on the logarithmic divergences. In fact, it is well known that scalar field tensor theories are not, in general, radiatively stable when coupled to dynamical matter. For example, consider the following proxy theory:
\begin{equation}
{\cal L} = -\frac{1}{2}(\partial\phi)^2 - \frac{1}{\Lambda_{\rm v}^3}\square\phi(\partial\phi)^2-\frac{1}{\Lambda_{\star}^4}(\partial\phi)^4 -\frac{1}{2}(\partial\chi)^2 -\frac{1}{2}M^2\chi^2 -\frac{g}{2\tilde{\Lambda}^2}(\partial\phi)^2 \chi^2\ ,
\end{equation}
where $\chi$ represents a matter field with mass $M$, $g$ measures the interaction strength between these two scalar sectors, and $\tilde{\Lambda}$ the corresponding strong coupling scale. Explicitly integrating out this heavy field, we find that the Euclidean action receives logarithmic contributions as follows:
\begin{equation}
	S_{\rm E}=\frac{1}{2}{\rm Tr\ log} \dfrac{\delta^2 S}{\delta \chi^2} \supseteq -\frac{g(\partial\phi)^2}{\tilde{\Lambda}^4} \left(
	M^2 \tilde{\Lambda}^2+\frac{g}{2} (\partial\phi)^2
	\right)\ {\rm{log}} \left(\frac{k_{\rm UV}^2+M^2}{\mu^2}\right),
\end{equation}
where we have explicitly ignored factors of order unity and introduced a sliding regularization scale, $\mu$.
These terms offer wavefunction renormalization to the classical low-energy field theory which are, by construction, large, in addition to a renormalization of the scale $\Lambda_{\star}$. Of course, which operators get renormalised will depend on the nature of the coupling to matter, which in the absence of an explicit UV completion, are unknown.

To bypass these difficulties one can invoke the Exact Renormalization Group~(ERG)~\cite{Wetterich:1992yh,Tetradis:1993ts,Rosten:2010vm}, which was applied in Ref.~\cite{deRham:2014wfa} to a simpler class of theories which also exhibit screening.
In this approach, a choice for the UV lagrangian is made---in this case for it to be of the form~\eqref{eq:galileonandgeneralpx}---and then the theory is run down to low energies, re-summing the loops involving all high-energy physics (both the heavy modes of the light field, $\phi$, as well as all the massive fields that $\phi$ couples to) along the way. 

Note that this approach also makes some optimistic assumptions about the UV physics, essentially because we posit a form of the UV lagrangian, but absent a UV completion, the level of tuning can not really be quantified. Solving the ERG equation is extremely hard, in particular because it includes not only contributions from the low-energy scalar, $\phi$, but also from other high-energy fields, which have been omitted in the EFT lagrangian, but one can gain useful insights even by solving it approximately. In this regime, one finds similar to~\cite{deRham:2014wfa} that corrections to the operators in~\eqref{eq:galileonandgeneralpx} are small so long as there is a large hierarchy between the scale of strong coupling and the true cutoff of the theory. That this could be the case in theories of this type has also recently been argued in~\cite{Keltner:2015xda}---see also Refs.~\cite{Dvali:2010jz,Dvali:2011nj} for analogous considerations in the context of classicalization.

\section{Conclusions}
\label{sec:summary}
In this paper we have considered the phenomenology of theories which are capable of exhibiting both kinetic and Vainshtein screening. Perhaps the most interesting possibility in these theories is that different screening mechanisms can be active for different mass objects or on different scales, which may allow for novel probes of these screening mechanisms. In particular it might be possible to evade solar system constraints via Vainshtein screening, but to have the (somewhat weaker) kinetic screening active on cluster scales.

We have focused on one concrete realization of this phenomenon---the galileon plus a $P(X)$ theory---because the phenomenology is  simple and it is straightforward to show that this corner of parameter space is radiatively stable. However, the same phenomenon should be exhibited by a broad subset of scalar-tensor theories, including the higher-order galileons, the conformal or DBI galileon~\cite{Nicolis:2008in,deRham:2010eu} and Horndeski~\cite{Horndeski:1974wa} or beyond Horndeski theories~\cite{Zumalacarregui:2013pma,Gleyzes:2014dya,Langlois:2015cwa}. The radiatively stable subset of~\cite{Pirtskhalava:2015nla} is particularly attractive in this regard. This generalization might make it easier to find completely satisfactory cosmological solutions. For the conformal galileon, an interesting feature of our analysis is that for sufficiently large mass sources to be screened the $(\partial\pi)^4$ term should have a negative sign---precisely the opposite from what is required in the proof of the $a$-theorem~\cite{Komargodski:2011vj}. This appears to be another avatar of the difficulties of UV completing theories which have screening mechanisms.

In the future it would also be interesting to understand what happens in theories which can exhibit both chameleon and kinetic/Vainshtein type screening; or even theories which exhibit all three. Further, it would be interesting to work out in detail the phenomenology of double screening for large scale structure, for example it is known that kinetic screening can be somewhat suppressed on quasi-linear scales~\cite{Brax:2014yla} and could allow for apparent violations of the equivalence principle~\cite{Creminelli:2013nua}. It would be useful to make quantitative estimates of these effects.

\paragraph{Acknowledgements:} We thank Kurt Hinterbichler, Bhuvnesh Jain and Justin Khoury for helpful discussions. RHR would like to thank the kind hospitality of the Kavli Institute for Cosmological Physics at the University of Chicago where this work was initiated. AJ \& RHR would like to kindly acknowledge the hospitality and vibrant atmosphere of the Sitka Sound Science Center in Alaska during the Sitka Sound Summer Workshop.
AJ \& WH thank the Aspen Center for Physics, which is supported by National Science Foundation grant PHY-1066293, where part of this work was performed. 
This work was also supported in part by the Kavli Institute for Cosmological Physics at the University of Chicago through grant NSF PHY-1125897, an endowment from the Kavli Foundation and its founder Fred Kavli, and by the Robert R. McCormick Postdoctoral Fellowship (AJ). RHR acknowledges support from the Science and Technology Facilities Council grant ST/J001546/1.  WH was supported 
additionally by U.S.~Dept.\ of Energy
contract DE-FG02-13ER41958 and NASA ATP NNX15AK22G.
\appendix

\section{Radiative stability of $G(X)\Box\phi$-type of theories}
\label{app:galileon}

Here we explore something somewhat outside the main line of development of this paper. We consider scalar-tensor theories which have a shift symmetry and take the form
\be
\mathcal{L}=-\dfrac{1}{2}(\partial\phi)^2+\Lambda G(X)\Box\phi \ ,
\label{eq:generalgalileon}
\ee
where $\Lambda$ is the strong coupling scale and $X\equiv -(\partial\phi)^2/2\Lambda^4$. These theories are often referred to as generalized galileons~\cite{Deffayet:2009mn} or Kinetic Gravity Braiding~\cite{Deffayet:2010qz}. In the main text, we chose to focus on the simplest case, $G(X) = X$, and instead allowed for kinetic nonlinearities of the $P(X)$ type.

Here we want to argue that choosing instead to consider a general $G(X)$ is not technically natural.  More precisely, we would like to know the full class of functions $G(X)$ for which the theory above is radiatively stable and can be consistently treated as a low-energy EFT. To address this question, we will again employ the heat kernel technique used in~Section~\ref{sec:rad}. To provide explicit results we focus on time-dependent background evolution for $\bar{\phi}$, but our results are otherwise generic. We find that, quantum-mechanically, the lowest-order generated operators take the following form: 
\be
\Delta\mathcal{L}\supset \dfrac{\ddot{\phi}^6}{\Lambda^{14}} \left\{
-2G_{,XX}-7XG_{,XXX}-2X^2G_{,XXXX}
\right\} \ .
\ee
On backgrounds where $\Box\phi\gtrsim\Lambda^3$, the operators above will be large. In this case, there is no consistent organization of the EFT. Therefore, we must choose for the terms between brackets to be small or vanish. This can of course be accomplished by choosing some specific functional form for $G$ such that it solves the differential equation $2G_{,XX}+7XG_{,XXX}+2X^2G_{,XXXX}\sim0$. However, there will be other radiatively generated operators beyond the lowest-combination above which will have a different operator structure, and which we will also have to demand be small---this would result in a very algebraically special theory which might not admit stable dynamics. We expect that generically, the only way to make all of these terms small is to take $G(X) = X$ so that they all vanish. It therefore seems that the only radiatively stable choice is the galileon. Note that this is in stark contrast to the situation for pure $P(X)$ theories, where essentially any functional form is radiatively stable with respect to logarithmic divergences~\cite{deRham:2014wfa}.

\bibliographystyle{utphys}
\addcontentsline{toc}{section}{References}
\renewcommand{\em}{}
\bibliography{screenbib}

\end{document}